\documentclass[runningheads]{llncs}
\usepackage{graphicx}
\usepackage{amsmath}
\usepackage{float}
\usepackage[caption=false]{subfig}
\usepackage{makecell}
\usepackage{comment}
\usepackage[title]{appendix}
\usepackage[linesnumbered, ruled]{algorithm2e}
\newtheorem{hyp}{Hypothesis}

\begin{document}
%
\title{New Quality Metrics for\\ Dynamic Graph Drawing\thanks{This work is supported by ARC DP grant.}}
%
%
\author{Amyra Meidiana \and
Seok-Hee Hong \and
Peter Eades}
\authorrunning{A. Meidiana et al.}

\institute{University of Sydney, Australia \\
\email{amei2916@uni.sydney.edu.au, \{seokhee.hong, peter.eades\}@sydney.edu.au}}
\maketitle              
\begin{abstract}
In this paper, we present new quality metrics for dynamic graph drawings.
Namely, we present a new framework for {\em change faithfulness} metrics for dynamic graph drawings, which compare the {\em ground truth change} in dynamic graphs and the {\em geometric change} in drawings.

More specifically, we present two specific instances, {\em cluster  change} faithfulness metrics and {\em distance change} faithfulness metrics.
We first validate the effectiveness of our new metrics using deformation 
experiments. 
Then we compare various graph drawing algorithms using our metrics. Our experiments confirm that the best cluster (resp. distance) faithful graph drawing algorithms are also cluster (resp. distance) change faithful.

\end{abstract}

\section{Introduction}

Quality metrics (or \textit{aesthetic criteria}~\cite{battista1998graph}) for graph drawings play an important role in evaluating graph drawings as well as designing new algorithms to optimize the metrics.
Traditional quality metrics for graph drawings mainly evaluate the {\em readability} of a drawing, such as edge crossings, edge bends, total edge length, and angular resolution~\cite{battista1998graph}. Most of these metrics focus on \textit{static} graphs. 

Network data are abundant in various domains, from social media to chemical pathways, and they are often changing with dynamics. Compared to static graph drawing, dynamic graph drawing brings its own challenges, such as the preservation of the user's mental map as the drawing evolves~\cite{eades1991preserving}. 
To evaluate dynamic graph drawing algorithms, we need quality metrics to measure how well a drawing of a dynamic graph reflects the changes in the graph.

\textit{Faithfulness} metrics measure how faithfully the ground truth about the data is displayed in the visualization~\cite{nguyen2013faithfulness}. 
For dynamic graphs, \textit{change faithfulness} measures how proportional the change in the drawings of dynamic graphs is to the change in the graphs.

However, existing work on quality metrics of dynamic graph drawings, such as preservation of the mental map~\cite{branke2001dynamic,diehl2002graphs,eades1991preserving}, mainly focus on the {\em readability} metrics, which only measure the geometric change in the drawing without considering how well the change represents the change in the graph. 
Furthermore, recent qualitative studies have shown that mental map preservation alone may not be sufficient to aid users in understanding dynamic graphs~\cite{archambault2013map}.

In this paper, we present a new framework for \textit{change faithfulness metrics} of dynamic graphs, quantitatively measuring how faithfully the ground truth change in dynamic graphs is proportionally displayed as the geometric change in dynamic graph drawings.

Based on the framework, we present two new quality metrics, {\em cluster  change} faithfulness metrics and {\em distance change} faithfulness metrics.
We validate the effectiveness of our new metrics using deformation 
experiments, and then compare various graph drawing algorithms using our metrics. 

More specifically, we present the following contributions:

\begin{enumerate}
\item We present a general \textit{change faithfulness metric} framework for dynamic graphs, which compares the ground truth change in dynamic graphs and the geometric change in the drawings.
\item We present the \textit{cluster change faithfulness metrics} $CCQ$ as an instance of the change faithfulness metrics, comparing the change in \textit{ground truth clustering} of dynamic graphs to the change in \textit{geometric clustering} of the drawing.
\item We present the \textit{distance change faithfulness metrics} $DCQ$ as another specific instance of the change faithfulness metrics, which compares the change in \textit{graph theoretic distance} of dynamic graphs to the change in \textit{geometric distance} of the drawing.
\item We validate the effectiveness of the cluster change faithfulness metrics and distance change faithfulness metrics using deformation experiments on drawings. 
Results of the experiments confirm that the $CCQ$ and $DCQ$ metrics decrease as the drawings are distorted such that the change between drawings are more disproportionate to the change in ground truth information.
\item We compare various graph drawing algorithms using the $CCQ$ and $DCQ$ metrics. Experiments confirm that the most cluster faithful layouts and distance faithful layouts indeed also obtain high cluster change faithfulness and distance faithfulness respectively. 
Interestingly, we also discover that in some cases, higher information faithfulness does not necessarily lead to higher change faithfulness.
\end{enumerate}

\section{Related Work}
\label{sec:litreview}

\subsection{Quality Metrics for Graph Drawing}

Traditional aesthetic criteria~\cite{battista1998graph} for graph drawings are mainly concerned with the {\em readability} of graphs, such as the minimization of edge crossings, bends, total edge lengths and drawing area. 
They have been established as criteria to be optimized by graph drawing algorithms~\cite{battista1998graph}. 

HCI studies have verified the correlation between aesthetic criteria with specific task performance on graphs. 
For example, few edge crossings~\cite{purchase1997aesthetic,purchase1995validating} and large crossing angles~\cite{huang2008effects} are important criteria for finding shortest paths between two vertices.
However, these studies tend to focus on {\em small} graphs.

More recently, a new concept of \textit{faithfulness} metrics have been introduced for {\em large} graphs, measuring how faithfully the ground truth information of graphs is displayed in graph drawings~\cite{nguyen2013faithfulness}. Subsequently, a series of new faithfulness metrics have been developed~\cite{eades2017shape,meidiana2019quality,meidiana2020quality,meidiana2019sym}.

\textit{Shape-based metrics}~\cite{eades2017shape} are introduced to evaluate {\em large} graph drawings, where traditional metrics such as edge crossings do not scale well. 
More specifically, the metrics compare the similarity between the original graph $G$ with a shape graph (or proximity graph) $G'$ computed from a  drawing $D$ of $G$. 

The \textit{cluster metrics} $CQ$~\cite{meidiana2019quality,meidiana2019sym} measure how faithfully the ground truth clusters of a graph is displayed in a drawing, by comparing the ground truth clusters to the geometric clustering in a graph drawing.

The \textit{symmetry metrics} ~\cite{meidiana2020quality} measure how faithfully the ground truth {\em automorphisms} of a graph (rotational or axial) and {\em automorphism groups} (cyclic or dihedral), are displayed as symmetries in a drawing, computed by approximate symmetry detection algorithms in $O(n \log n)$ time.
A $O(n \log n)$ time algorithm for exact symmetry detection is also presented.

\subsection{Quality Metrics for Dynamic Graph Drawing}

A \textit{dynamic} graph is defined by a sequence of static graphs $G_1, G_2, \ldots, G_k$ spanning $k$ time steps, where $G_i$ is a time slice of the graph at time step $i$~\cite{beck2017taxonomy}. 
Dynamic graphs are most commonly visualized using small multiples~\cite{tufte1990envisioning} or animation.

A long standing challenge with dynamic graph drawings is preserving the user's \textit{mental map}~\cite{eades1991preserving}, where dramatic changes in the positions of vertices can make it difficult for users to keep track of the state of a dynamic graph. 
The mental map can be modelled using e.g. orthogonal ordering, clustering, or topology~\cite{eades1991preserving}. Related is the concept of \textit{dynamic stability}, which aims to minimize the geometric distance between subsequent drawings~\cite{bohringer1990using,tamassia1988automatic}.  Stability has been shown to assist users in performing analytical tasks on dynamic graphs~\cite{archambault2016can}.

A recent survey on dynamic graph drawing~\cite{beck2017taxonomy} addresses that evaluation is one of the most important research questions on dynamic graph drawings. 
Quantitatively, dynamic graph drawings can be evaluated using \textit{distance} metrics, including Euclidean distance, orthogonal distance, and edge routing, to measure the extent of mental map preservation~\cite{branke2001dynamic,diehl2002graphs}. 

However, specific change faithfulness metrics for dynamic graph drawings have yet to be developed to measure how the ground truth change in dynamic graphs are proportionally displayed as geometric change in drawings.

\section{Change Faithfulness Metric Framework}
\label{sec:changemetric}

\begin{figure}[t!]
\centering
\includegraphics[width=0.8\textwidth]{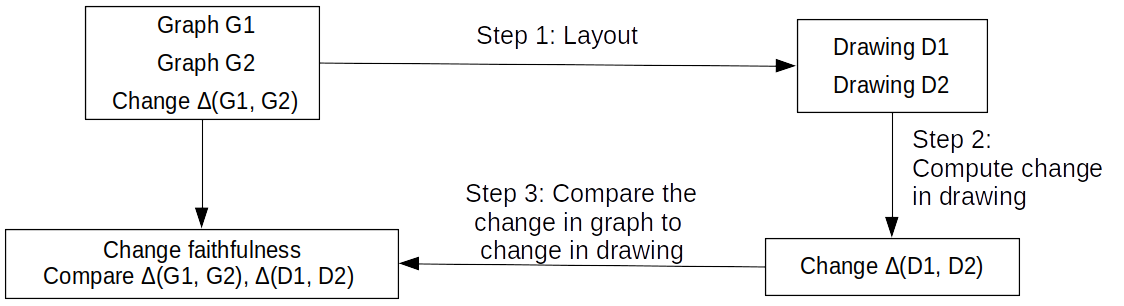}
\caption{Change faithfulness metric framework:   
The change faithfulness metric is computed by comparing the ground truth change $\Delta(G_1, G_2)$ between two graphs $G_1$ and $G_2$, and the geometric change $\Delta(D_1, D_2)$ in drawings of graphs. }
\label{fig:changeframework}
\end{figure}

We propose the \textit{change faithfulness metric} for measuring how well dynamic graph drawings show the structural changes in dynamic graphs. 
Roughly speaking, a drawing is \textit{change faithful}
if the extent of change in the drawing is proportional to the extent of (ground truth) change in the graph. 
Fig.~\ref{fig:changeframework} illustrates the general framework for change faithfulness metrics.

In practice, the vertex set of a dynamic graph may change; in this paper we focus on cases where only the edge set changes. 
Let $G_1 = (V,E_1)$ and $G_2 = (V,E_2)$ be two time slices of a dynamic graph, with the change denoted as $\Delta(G_1, G_2)$.
The change faithfulness metrics are computed as follows:

\begin{description}
\item[Step 1:] Compute a drawing $D_1$ (resp. $D_2$) of $G_1$ (resp. $G_2$).

\item[Step 2:]  Compute the geometric change $\Delta(D_1, D_2)$ between $D_1$ and $D_2$.

\item[Step 3:] Compute the change faithfulness metrics by comparing the ground truth change $\Delta(G_1, G_2)$ to $\Delta(D_1, D_2)$.
\end{description}

The framework in 
Fig.~\ref{fig:changeframework} is a general framework applicable to various types of change in dynamic graphs. 
The detailed definitions for $\Delta(G_1, G_2)$ and $\Delta(D_1, D_2)$, as well as how to compare them, depend on the nature of the considered change.

\subsection{Cluster Change Faithfulness Metrics}

We present the \textit{cluster change faithfulness metric} $CCQ$ as an example of a change faithfulness metric. 
$CCQ$ measures how faithfully the change in \textit{ground truth clustering} is reflected as a change in the \textit{geometric clustering} between drawings of different time slices of a dynamic graph. Figure \ref{fig:clustchangeframework} in Appendix \ref{sec:ccqframework} summarizes the framework.

Let $C_1$ (resp. $C_2$) be the ground truth clustering of the vertices of $G_1$ (resp. $G_2$), with the change between the clusterings denoted as $\Delta(C_1,C_2)$. 
The cluster change faithfulness is defined as follows:

\begin{description}
\item[Step 1:] Compute a drawing $D_1$ (resp. $D_2$) of $G_1$ (resp. $G_2$).

\item[Step 2:] Compute geometric clustering $C'_1$ (resp. $C'_2$) based on vertex positions in $D_1$ (resp. $D_2$), and compute the change in geometric clustering $\Delta(C'_1,C'_2)$.

\item[Step 3:] Compute $CCQ$ by comparing $\Delta(C_1,C_2)$ to $\Delta(C'_1,C'_2)$.
\end{description}

To compute $\Delta(C_1,C_2)$ and $\Delta(C'_1,C'_2)$, any clustering comparison metrics can be used. 
In this paper, we use $ARI$ (Adjusted Rand Index)~\cite{hubert1985comparing,rand1971objective} and $FMI$ (Fowlkes-Mallows Index)~\cite{fowlkes1983method},
which showed superior performance in  measuring cluster faithfulness in static graph drawing~\cite{meidiana2019quality}.
$ARI$ is based on the number of pairs of elements classified into the same and different groups in two clusterings of the same set.  
$FMI$ is computed using the number of true positives, false positives, and false negatives. 

For Step 2, any geometric clustering algorithm can be used to compute $C'_1$ and $C'_2$. 
In this paper we use \textit{k-means clustering}, which partitions a set into \(k\) subsets that minimize the within-class variance~\cite{macqueen1967some}. 
It is a widely used clustering method with efficient heuristic approximation. 

For Step 3, we define $CCQ$ as follows:
\begin{equation}
    CCQ = 1 - \frac{|\Delta(C_1,C_2)-\Delta(C'_1,C'_2)|}{max(\Delta(C_1,C_2),\Delta(C'_1,C'_2))}
    \label{eq:ccq}
\end{equation}

Specifically, we take the difference between $\Delta(C_1,C_2)$ to $\Delta(C'_1,C'_2)$, and express the difference as a fraction of the larger value between the two, as both are normalized to the same range by using the same clustering quality metrics. 
We then negate the result from 1, such that 1 represents completely change faithful drawings and less change faithful drawings obtain values closer to 0.

\subsection{Distance Change Faithfulness Metrics}

We also present the \textit{distance change faithfulness metric} $DCQ$ as another instance of change faithfulness metric. 
We define \textit{distance faithfulness} as how faithfully graph theoretic distances between vertices in a graph are displayed as geometric distances between the positions of vertices in a drawing. 
Similarly, \textit{distance change faithfulness} measures how faithfully the change in graph theoretic distances is reflected as a proportional change in the geometric distances.
Figure \ref{fig:distchangeframework} in Appendix \ref{sec:dcqframework} presents the framework.

Let $\Delta(SP_1,SP_2)$ be the change in graph theoretic distances between two time slices of a dynamic graph, $G_1$ and $G_2$. More specifically, the distance change faithfulness metric is defined as follows:

\begin{description}
\item[Step 1:] Compute a drawing $D_1$ (resp. $D_2$) of $G_1$ (resp. $G_2$).

\item[Step 2:] Compute the change in geometric distance $\Delta(GD_1,GD_2)$.

\item[Step 3:] Compute $DCQ$ by comparing $\Delta(SP_1,SP_2)$ to $\Delta(GD_1,GD_2)$.

\end{description}

One example measure of distance faithfulness is \textit{stress}~\cite{battista1998graph}.
For each pair of vertices $v_i, v_j$ in a graph $G$, the stress is defined by the difference between the graph theoretic distance (i.e., shortest path) between $v_i$ and $v_j$, and the geometric distance between the positions of $v_i$ and $v_j$ in a drawing $D$ of $G$.

Using stress measures, we present two types of distance change faithfulness metrics $DCQ$. 
The first metric $DCQ_1$ is based on the \textit{target edge length} used in some stress-based layouts (e.g.~\cite{gansner2012maxent}). 
Given a target edge length $tl$, we expect neighboring vertices (i.e. path length 1) to have a geometric distance of $tl$. We thus scale the geometric distance between each pair of vertices in $D$ by $tl$. 

Let $\Delta(v_i,v_j) = |\delta_1(v_i,v_j)-\delta_2(v_i,v_j)|/max(\delta_1(v_i,v_j),\delta_2(v_i,v_j))$ and $S(v_i,v_j)=|s_1(v_i,v_j)-s_2(v_i,v_j)|/max(s_1(v_i,v_j),s_2(v_i,v_j))$,  where $\delta_1(v_i,v_j)$ (resp. $\delta_2(v_i,v_j)$) is the graph theoretic distance between vertices $v_i,v_j$ in $G_1$ (resp. $G_2$) and $s_1(v_i,v_j)$ (resp. $s_2(v_i,v_j)$) is the geometric distance between vertices $v_i,v_j$ in $D_1$ (resp. $D_2$). 
Scaling $S(v_i,v_j)$ by $tl$ to ensure the change in geometric distance is scaled to the target edge length, we define $DCQ_1$ as follows:
\begin{equation}
    DCQ_1 = 1 - \frac{2}{|V|^2}\sum^{|V|}_{i = 0}\sum^{|V|}_{j=i+1} \left| \Delta(v_i,v_j) - \frac{S(v_i,v_j)}{tl} \right|
    \label{eq:dcq1}
\end{equation}

In practice, not every layout algorithm takes an target edge length as input. 
Therefore, we instead use the \textit{average} of all edge lengths as $tl$.

For the second type of distance change faithfulness metric $DCQ_2$, we scale both the graph theoretic and geometric distances by the \textit{maximum} distance. 
For graph theoretic distances, it is the \textit{diameter} of graph $G$, while for geometric distances, it is the largest distance between any pair of vertices in drawing $D$. 

The scaled graph theoretic distance is given as $\delta'(i,j) = \delta(v_i,v_j)/diam(G)$, where $diam(G)$ is the diameter of $G$. 
The scaled geometric distance is given as $s'(i,j) = s(v_i,v_j)/max(s)$, where $max(s)$ is the maximum distance between any two vertices in $D$. 
We define $DCQ_2$ as follows:
\begin{equation}
    DCQ_2 = 1 - \frac{2}{|V|^2}\sum^{|V|}_{i = 0}\sum^{|V|}_{j=i+1}  \left| |\delta'_1(i,j)-\delta'_2(i,j)| - |s'_1(i,j)-s'_2(i,j)|\right|
    \label{eq:dcq2}
\end{equation}

\section{Cluster Change Faithfulness Validation Experiment}
\label{sec:clustchangeval}


To validate the cluster change faithfulness metrics, we design {\em deformation} experiments. 
Given two dynamic graph time slices $G_1$ and $G_2$ with ground truth clustering $C_1$ and $C_2$, we start with \textit{cluster faithful drawings} $D_1$ and $D_2$, i.e. the geometric clustering $C'_1$ of $D_1$ (resp. $C'_2$ of $D_2$) is the same as $C_1$ (resp. $C_2$). This gives $\Delta(C_1, C_2) = \Delta(C'_1,C'_2)$, i.e. cluster change faithful.

We then progressively deform drawing $D_2$. In each experiment, we perform 10 steps of deformation, where in each step, the coordinates of each vertex from the previous step are perturbed by a value in the range \([0,\delta]\), where \(\delta\) is the size of the drawing area multiplied by a value in the range [0.05,0.1]. 
We compute $CCQ$ and compare the scores across all steps of the deformation.

We expect that $CCQ$ will decrease with the deformation steps, as $\Delta(C'_1,C'_2)$ will grow further away from $\Delta(C_1, C_2)$. 
We formulate the following hypothesis:
\begin{hyp}\label{hyp:clustchangeval}
    $CCQ_{ARI}$ and $CCQ_{FMI}$ decrease as $D_2$ is deformed.
\end{hyp}

We generate ten dynamic graph data sets for the $CCQ$ validation experiment, with 200-1000 vertices each, as follows: First, we create a small graph (up to 30 vertices). We replace each vertex with a larger, denser graph, which becomes a cluster in $G_1$. 
We then replace each edge with inter-cluster edges between a randomly selected subset of vertices from each cluster. 
To create $G_2$, we change the cluster membership of vertices, either by merging clusters through randomly adding inter-cluster edges until a desired density for the new cluster is achieved, or splitting clusters by deleting edges between two partitions of the cluster until a desired lower intra-cluster edge density is reached.

To compute the initial layouts, we use the Backbone layout from Visone~\cite{baur2001visone}, which produces cluster faithful layouts (i.e. $CQ=1$) for our validation datasets. 
We use cluster comparison metric implementation from scikit-learn~\cite{pedregosa2011scikit}.

\begin{figure}[t!]
\centering
\subfloat[$D_1$]{
\includegraphics[width=0.22\columnwidth]{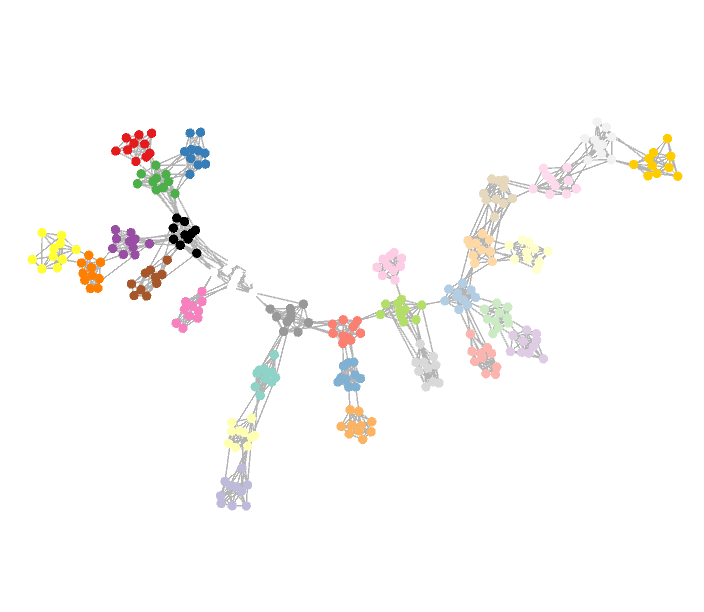}
}
\subfloat[$D_2$ step 0]{
\includegraphics[width=0.22\columnwidth]{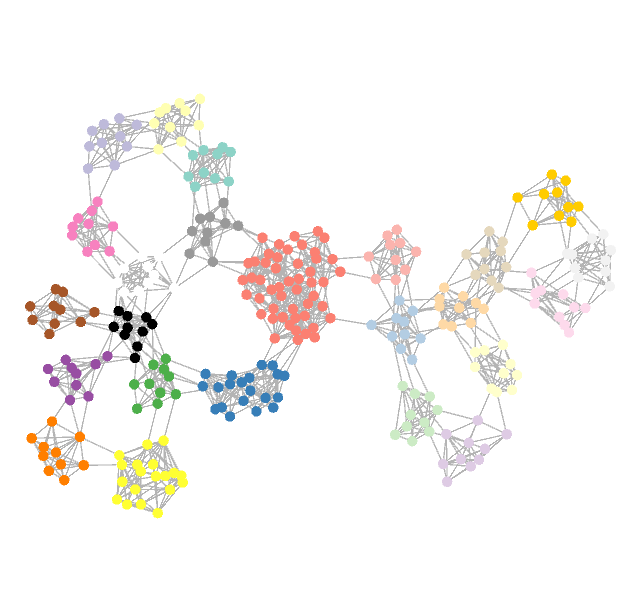}
}
\subfloat[$D_2$ step 3]{
\includegraphics[width=0.22\columnwidth]{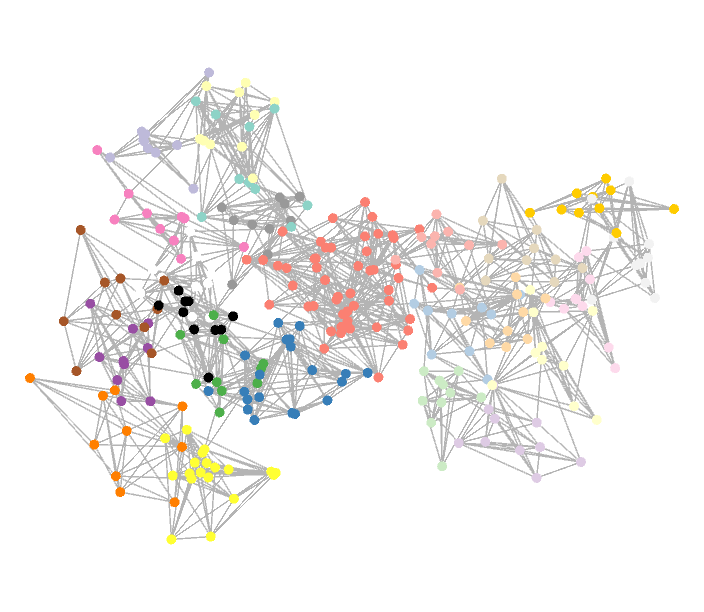}
}
\subfloat[$D_2$ step 10]{
\includegraphics[width=0.22\columnwidth]{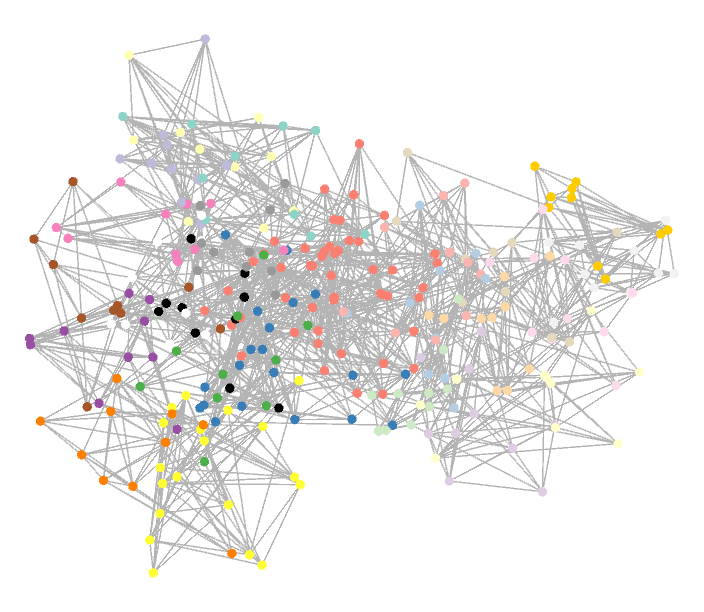}
}
\caption{Deformation experiment for $clusts-tree-30$,  showing deformation steps.}
\label{fig:perturb-clust}
\end{figure}

Fig. \ref{fig:perturb-clust} shows a deformation experiment example, where vertices are colored based on ground truth cluster membership. 
Figs. \ref{fig:perturb-clust} (a) and (b) show $D_1$ and $D_2$ at step 0. 
As the positions are perturbed in Figs. \ref{fig:perturb-clust} (c) and (d), vertices in the same cluster grow further apart and mix with vertices from other clusters, making the drawing less cluster faithful and subsequently increasing the difference between the geometric clustering in $D_1$ and $D_2$.

\begin{figure}[t]
    \centering
    \includegraphics[width=0.6\textwidth]{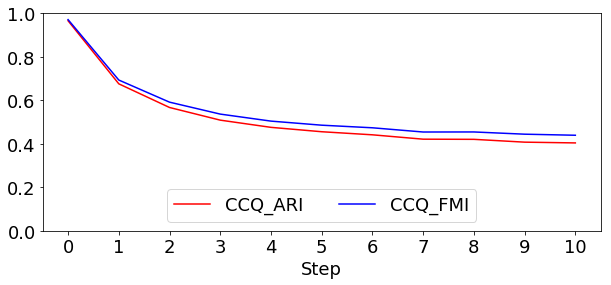}
    \caption{Average of $CCQ$ for all validation experiments. The decreasing trend for all versions of $CCQ$ validates Hypothesis \ref{hyp:clustchangeval}.}
    \label{fig:clustchange_perturb_average}
\end{figure}

Fig. \ref{fig:clustchange_perturb_average} shows the average $CCQ$ scores for each deformation step, averaged for all data sets. 
Clearly, we can see that $CCQ$ metrics decrease after each deformation step, validating Hypothesis \ref{hyp:clustchangeval}.

\subsection{Discussion and Summary}

Fig. \ref{fig:clustchange_perturb_average} clearly shows a downward slope of the $CCQ$ metrics,
which validates the usage of both $CCQ_{ARI}$ and $CCQ_{FMI}$ metrics  with our framework. 
Previous work on cluster faithfulness metrics $CQ$ on static graphs~\cite{meidiana2019quality} has shown that $ARI$ is more sensitive to changes than $FMI$. 
To a lesser extent, a similar pattern can be seen here, where $CCQ_{ARI}$ decreases to a lower score on latter perturbation steps compared to $CCQ_{FMI}$, indicating that it may be better in capturing changes in cluster change faithfulness as well.

\textit{In summary, the validation experiments have shown that the $CCQ$ metrics effectively reflect the cluster change faithfulness of drawings of dynamic graphs with dynamic clusters. 
Furthermore, we see that $CCQ_{ARI}$ is slightly more effective in capturing cluster change faithfulness than $CCQ_{FMI}$.}

\section{Cluster Change Faithfulness Layout Comparison}
\label{sec:clustlayoutcomp}

After validating the effectiveness of the cluster change faithfulness metrics, 
we use the $CCQ$ metrics to compare the performance of various graph drawing algorithms. 
We select the following layout algorithms: \textit{LinLog}~\cite{noack2003energy}, a force-directed layout emphasizing clusters; \textit{Backbone}~\cite{nocaj2015untangling}, which uses Simmelian backbones to extract communities; \textit{tsNET}~\cite{kruiger2017graph}, which uses t-SNE~\cite{maaten2008visualizing} and aims to preserve point neighborhoods; and \textit{sfdp}~\cite{hu2005efficient}, a multi-level force-directed layout.

LinLog, Backbone, and tsNET are designed to display clusters, and sfdp was seen to be more cluster faithful than other non-cluster-focused layouts~\cite{meidiana2019quality}. 

As LinLog was shown to be the most cluster faithful~\cite{meidiana2019quality}, we also expect it to be the most cluster change faithful, formulating the following hypothesis:

\begin{hyp}\label{hyp:clustchangecomp}
    LinLog scores the highest  $CCQ$ metrics.
\end{hyp}

\begin{table}[ht!]
\centering
\caption{Layout comparison on  \(gnm\_10\_25\)}
\label{table:clustlayoutcomp_gnm_10_25}
\begin{tabular}{|c|c|c|c|}
\hline
$G_1$ Backbone & $G_1$ LinLog & $G_1$ sfdp & $G_1$ tsNET \\ \hline
\includegraphics[width=0.22\textwidth]{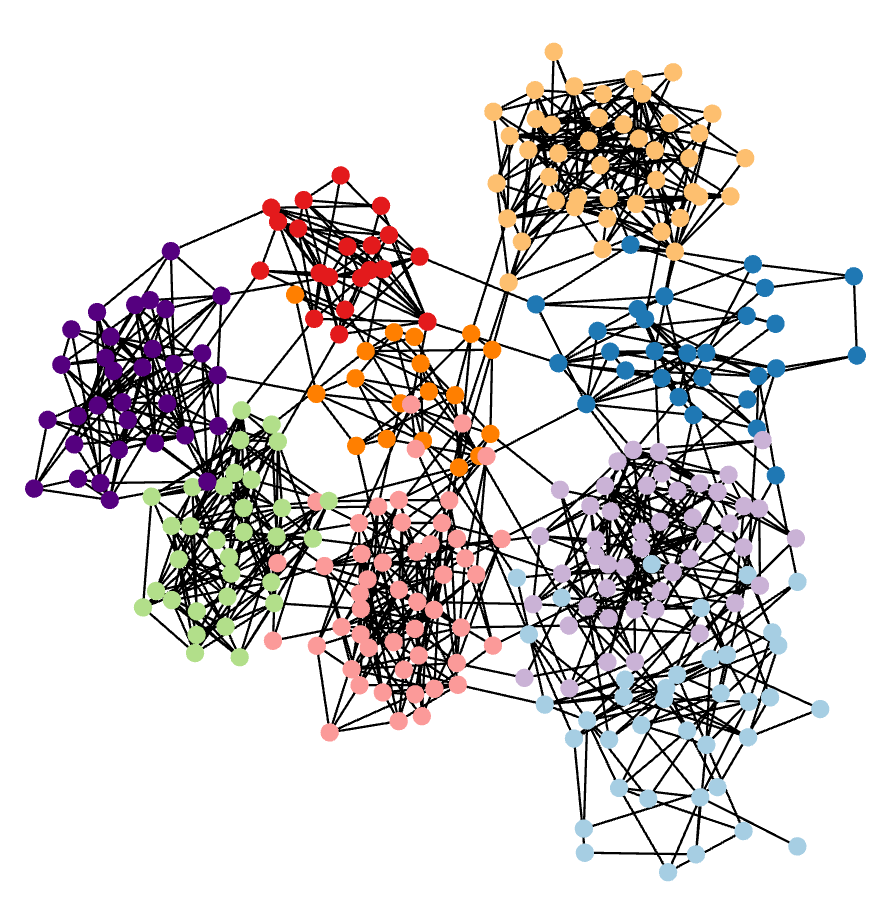} &
\includegraphics[width=0.22\textwidth]{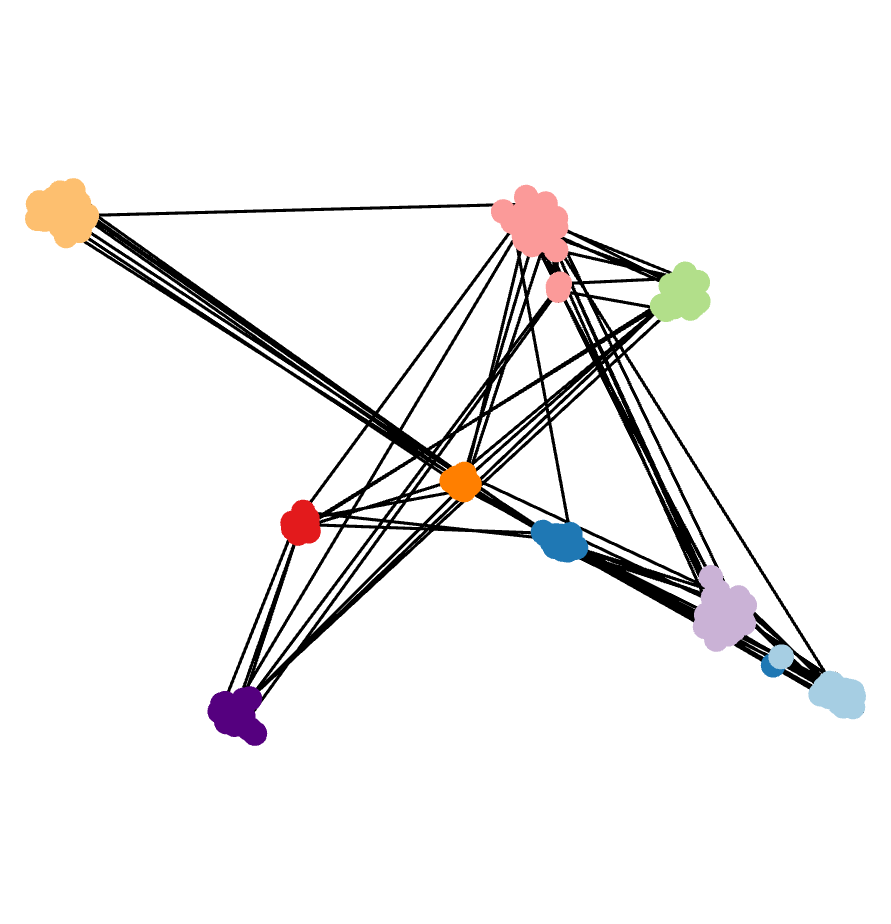} &
\includegraphics[width=0.22\textwidth]{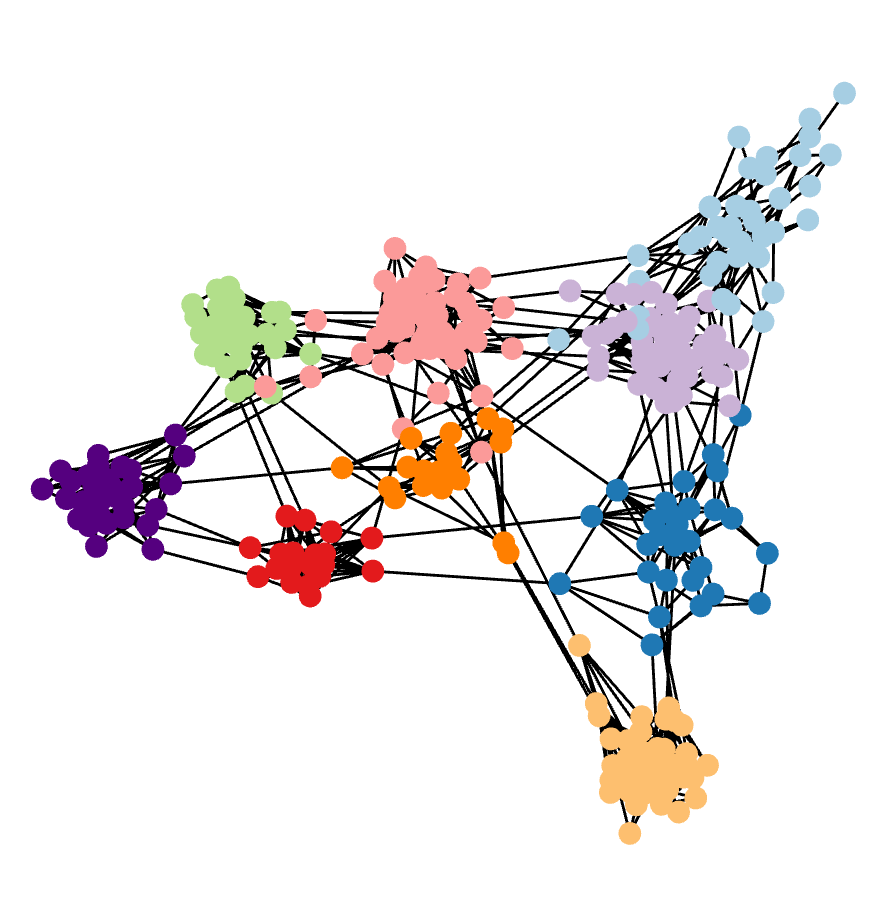} &
\includegraphics[width=0.22\textwidth]{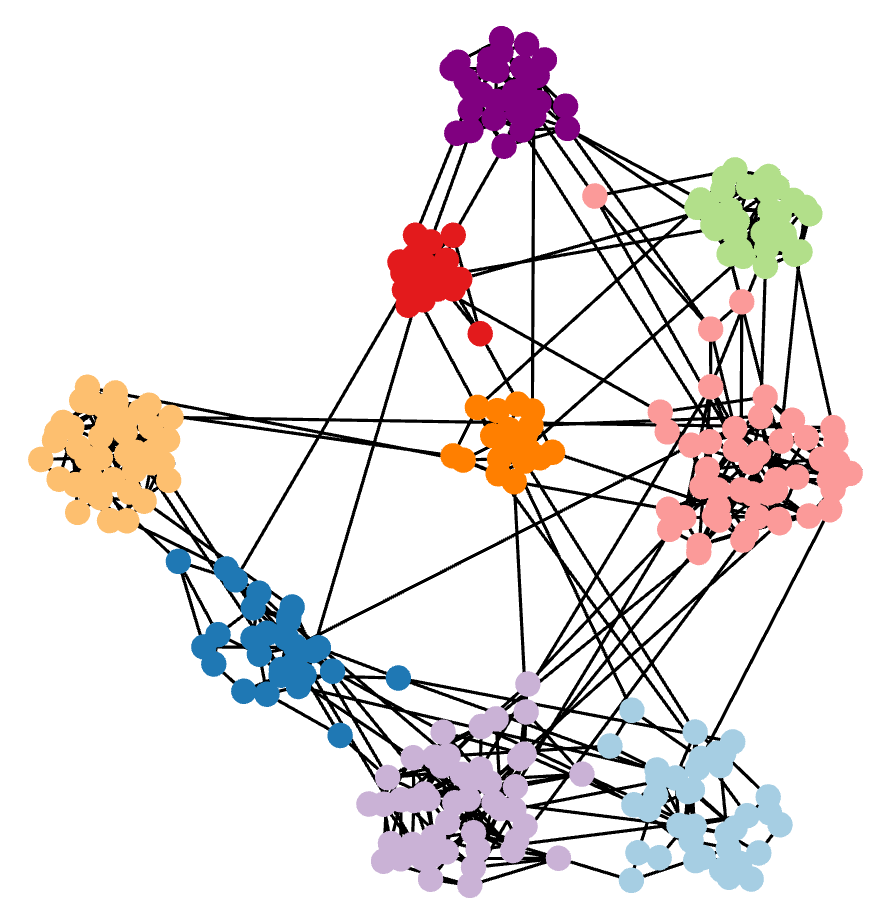}\\ \hline
$G_2$ Backbone & $G_2$ LinLog & $G_2$ sfdp & $G_2$ tsNET \\ \hline
\includegraphics[width=0.22\textwidth]{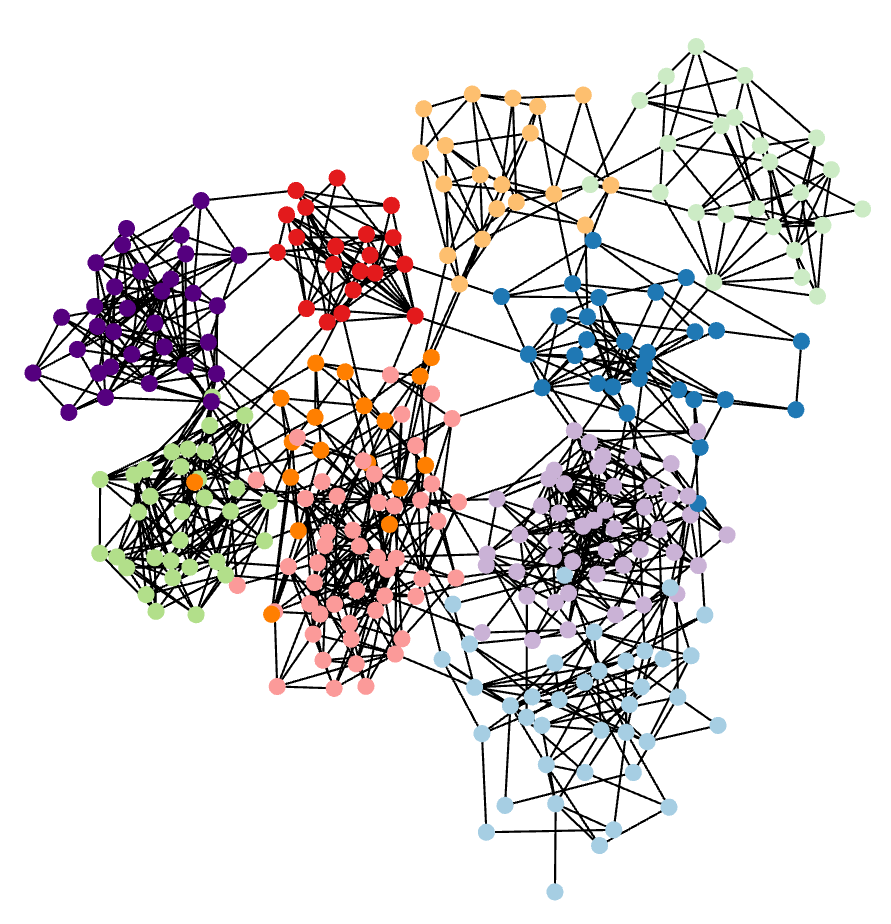} &
\includegraphics[width=0.22\textwidth]{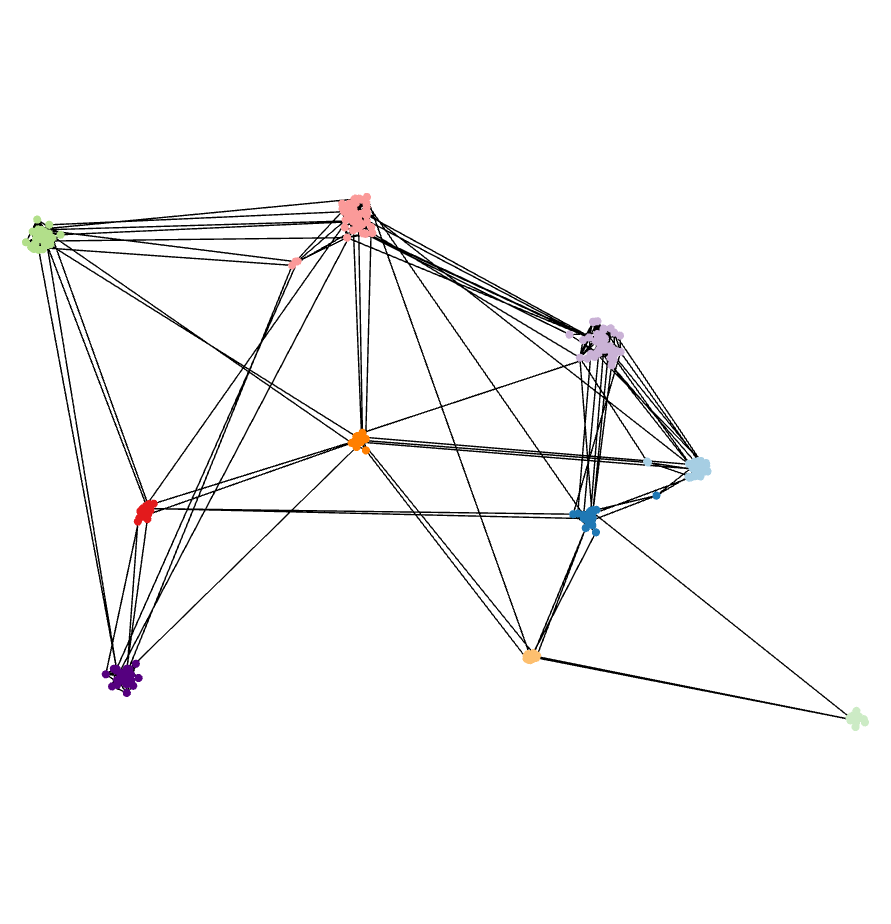} &
\includegraphics[width=0.22\textwidth]{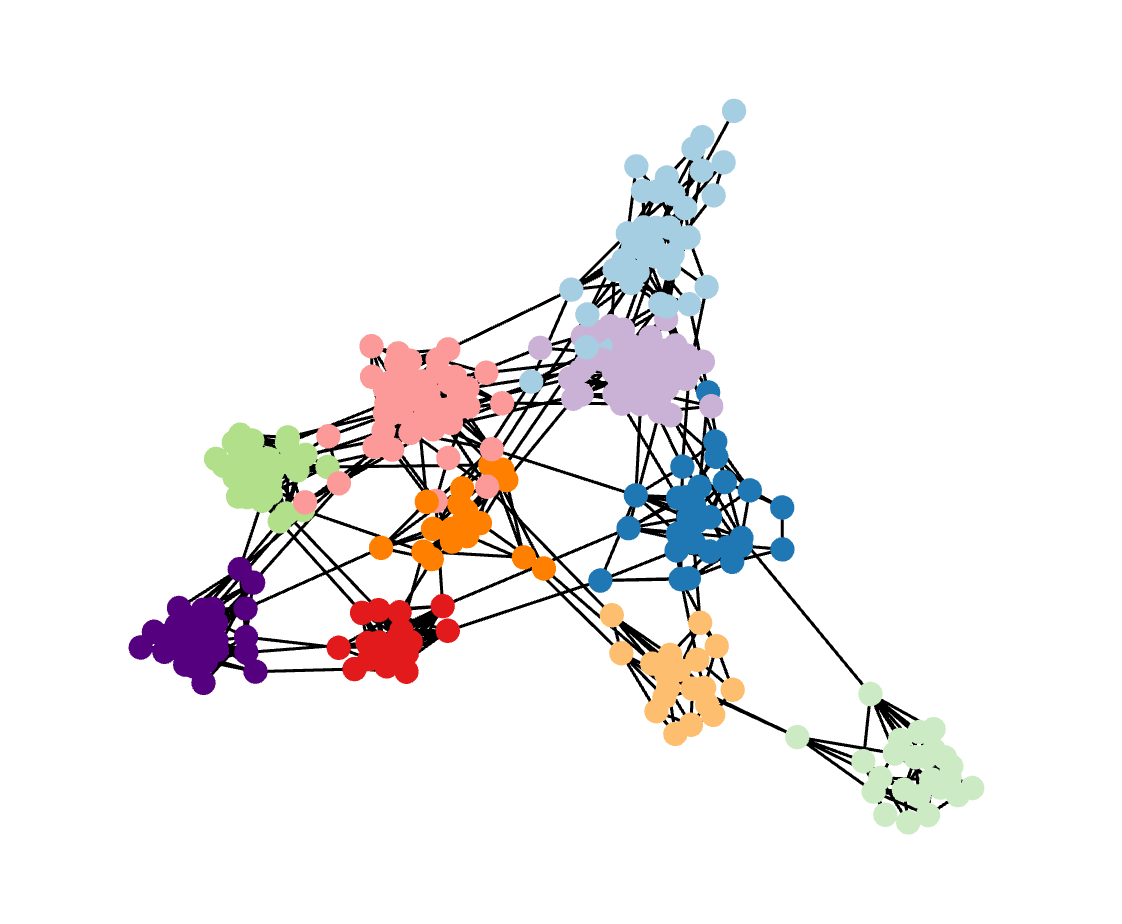} &
\includegraphics[width=0.22\textwidth]{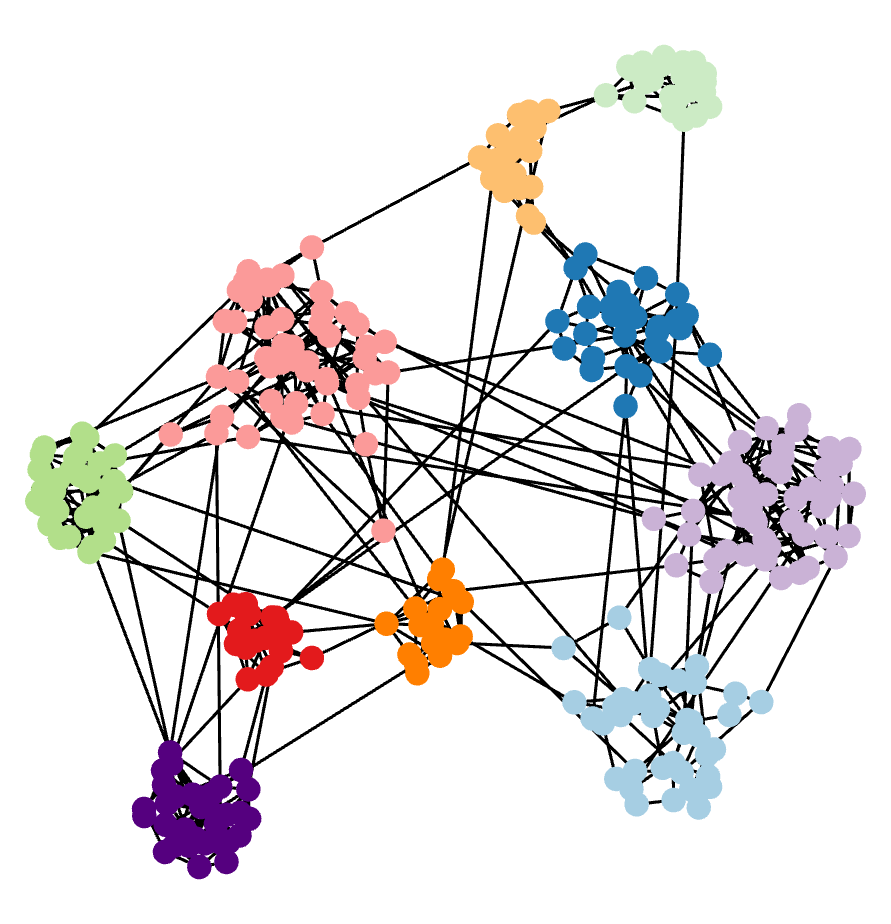}\\ \hline
\multicolumn{2}{|c|}{$CQ$} & \multicolumn{2}{c|}{$CCQ$} \\ \hline
\multicolumn{2}{|c|}{\includegraphics[width=0.4\columnwidth]{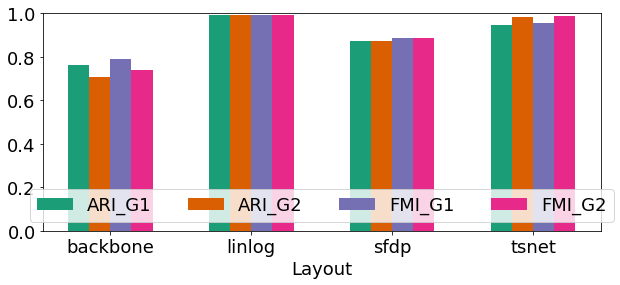}} & \multicolumn{2}{c|}{\includegraphics[width=0.4\columnwidth]{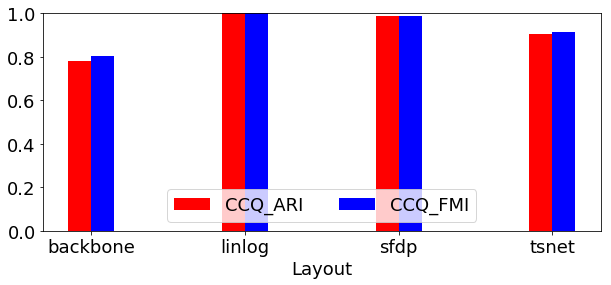}} \\ \hline
\end{tabular}
\end{table}

We use Tulip~\cite{david2001tulip} (LinLog), visone~\cite{baur2001visone} (Backbone), Graphviz~\cite{ellson2001graphviz} (sfdp), and  tsNET~\cite{kruiger2017tsnet}. 
We use thirteen dynamic graphs including synthetic data created similarly as in Section \ref{sec:clustchangeval}, and real-world data the Social Evolution set~\cite{madan2011sensing}; the graph sizes range from around 80-1000 vertices.


Table \ref{table:clustlayoutcomp_gnm_10_25} shows a layout comparison example, with a cluster split (yellow into yellow and pastel green). 
The $CQ$ cell shows the cluster faithfulness metrics: 
green and orange show the $CQ_{ARI}$ metric for $G_1$ and $G_2$ respectively, 
and purple and pink show the $CQ_{FMI}$ metric for $G_1$ and $G_2$ respectively. 
The $CCQ$ cell shows the $CCQ$ metrics: red for $CCQ_{ARI}$ and blue for $CCQ_{FMI}$.

LinLog obtains the highest $CCQ$ score, supporting Hypothesis \ref{hyp:clustchangecomp}. 
We also see a case of higher $CQ$ not always corresponding to higher $CCQ$: for example, tsNET obtains higher $CQ$ than sfdp, however, it obtains lower $CCQ$ than sfdp.

\begin{figure}[ht]
    \centering
    \includegraphics[width=0.6\textwidth]{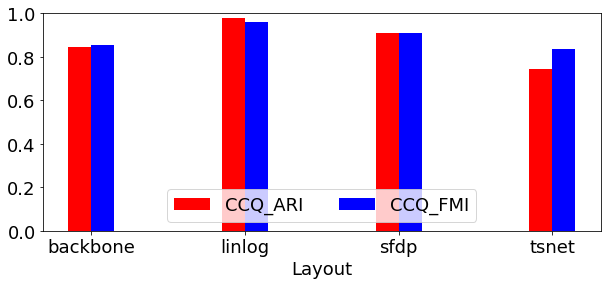}
    \caption{Average of $CCQ$ for layout comparison experiments. LinLog obtains the highest $CCQ$, validating Hypothesis \ref{hyp:clustchangecomp}. sfdp unexpectedly obtains the second highest $CCQ$.}
    \label{fig:clustchange_layoutcomp_avg}
\end{figure}

Fig. \ref{fig:clustchange_layoutcomp_avg} shows the average $CCQ$ scores across all data sets used for the layout comparison experiment. On average, LinLog obtains the highest $CCQ$ metrics, at 0.98 on $CCQ_{ARI}$, validating Hypothesis \ref{hyp:clustchangecomp}.

\subsection{Discussion and Summary}

Our experiments confirm that the LinLog layout, which was previously shown as the most cluster faithful layout for static graphs, also obtains the highest cluster change faithfulness for dynamic graphs.

We also find cases where better cluster faithfulness does not always correspond to better cluster change faithfulness, as seen in Table \ref{table:clustlayoutcomp_gnm_10_25}. 
This may be due to the clusters ``moving around'' between the drawings produced by tsNET, causing different misclassifications. 
For example, in $D_1$, some members of the pink cluster were misclassified to the dark purple or lime green clusters in $D_1$;  however, they are misclassified into the lime green or orange instead in $D_2$.

Meanwhile, sfdp produces drawings where relative positions of the cluster are more stable, causing the misclassifications to be more ``consistent'', e.g. members of the pink cluster are misclassified only into the lime green and orange clusters in both $D_1$ and $D_2$. Stability alone does not always lead to high change faithfulness, however, as seen from Backbone in Fig. \ref{table:clustlayoutcomp_gnm_10_25}, where the cluster positions are stable yet $CCQ$ is still low as $CQ$ is lower compared to the other layouts.

\textit{In summary, our experiments confirm Hypothesis \ref{hyp:clustchangecomp}, showing that LinLog produces the most  cluster change faithful drawings. We also show that cluster faithfulness does not always translate to cluster change faithfulness, in cases where subsequent drawings do not preserve the relative locations of the clusters.}

\section{Distance Change Faithfulness Validation Experiment}

We also validate the distance change faithfulness metrics, using validation experiments. 
Given two graph time slices $G_1$ and $G_2$, we start with \textit{stress faithful drawings} $D_1$ and $D_2$. 
We then perturb $D_2$ as follows: before perturbing, we divide the edges into two sets $E'_1$ and $E'_2$. In each step, we select edges from $E'_1$ to extend their lengths, and select edges from $E'_2$ to shorten their lengths.

We expect that the $DCQ$ scores  decrease with the deformation steps. We therefore formulate the following hypothesis:

\begin{hyp}\label{hyp:distchangeval}
    $DCQ_1$ and $DCQ_2$  decrease as the drawing $D_2$ is deformed, and $DCQ_1$ performs  better than $DCQ_2$ in  measuring distance change faithfulness.
\end{hyp}

To create the validation data sets, we start with a randomly-generated graph $G_1$, typically with a long diameter. To create $G_2$, we add edges to $G_1$ that significantly reduces the diameter and introduces smaller cycles into the graph. 
We generate ten dynamic graphs with 20-300 vertices and draw them using the \textit{Stress Majorization} layout from Tulip~\cite{david2001tulip} to obtain low stress drawings.


\begin{figure}[t]
\centering
\subfloat[$D_1$]{
\includegraphics[width=0.22\columnwidth]{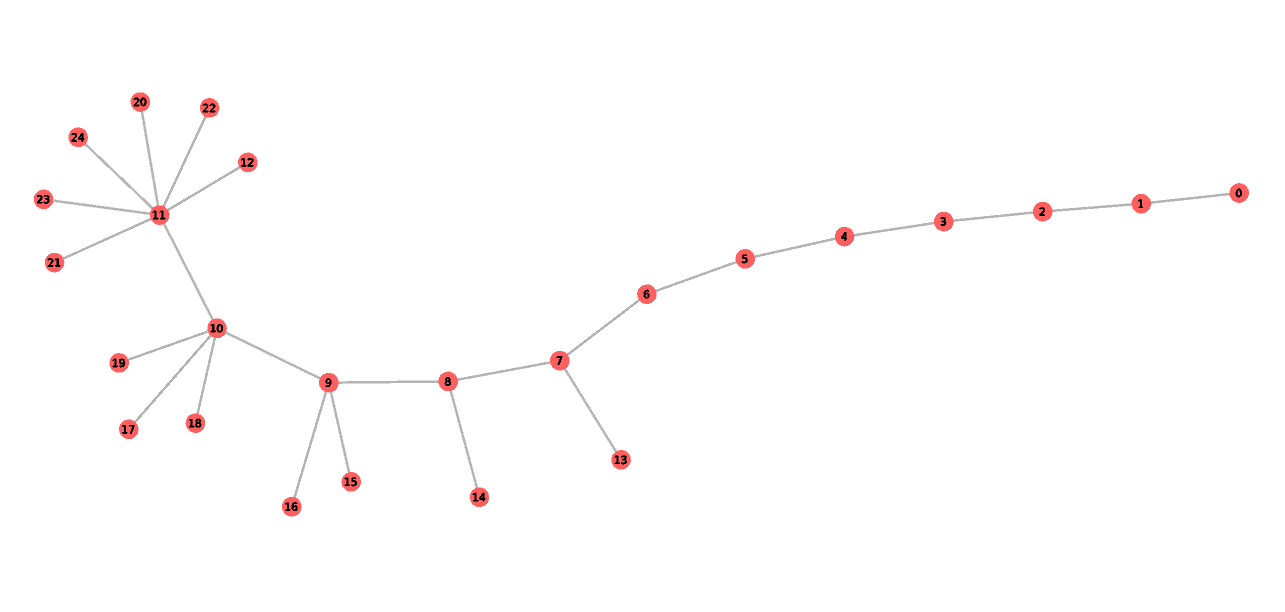}
}
\subfloat[$D_2$ step 0]{
\includegraphics[width=0.22\columnwidth]{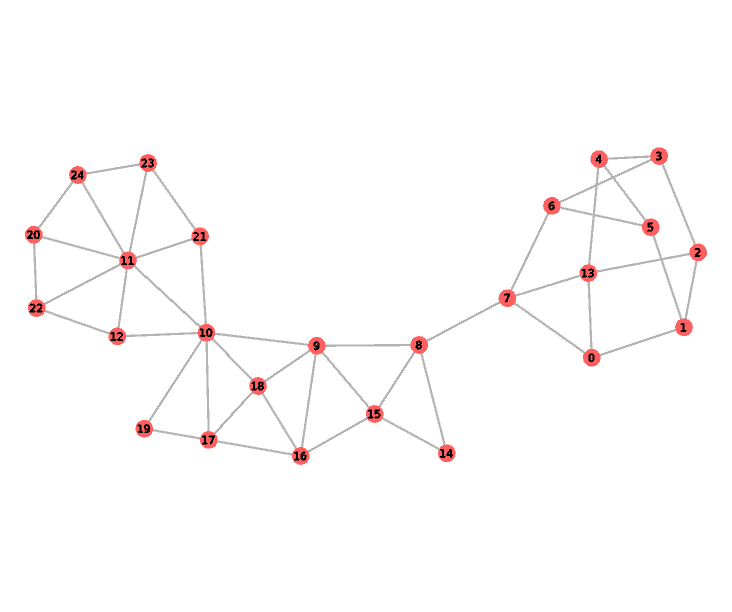}
}
\subfloat[$D_2$ step 3]{
\includegraphics[width=0.22\columnwidth]{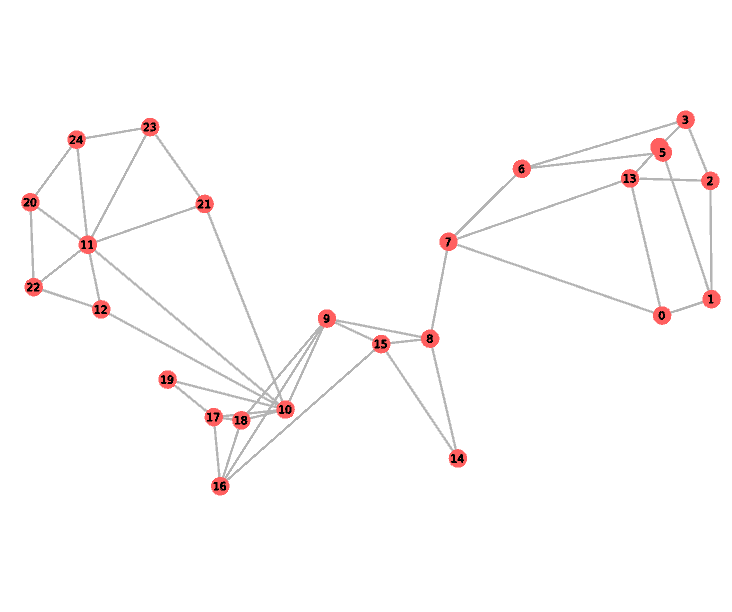}
}
\subfloat[$D_2$ step 10]{
\includegraphics[width=0.22\columnwidth]{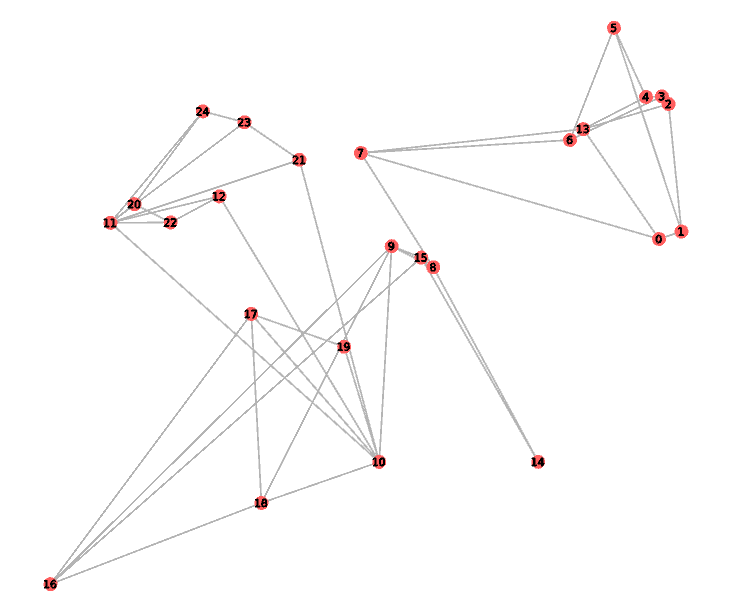}
}
\caption{Deformation experiment for $powertree\_25\_1$,
showing deformation steps.}
\label{fig:perturb-dist}
\end{figure}

\begin{figure}[t]
    \centering
    \includegraphics[width=0.6\textwidth]{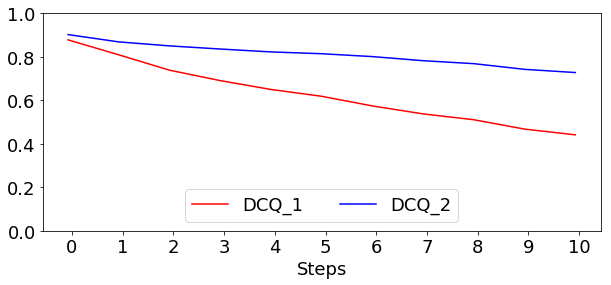}
    \caption{Average of $DCQ$ metrics for all validation experiments. The decreasing trend, especially with $DCQ_1$, validates Hypothesis \ref{hyp:distchangeval}.}
    \label{fig:distchange_perturb_average}
\end{figure}

Fig. \ref{fig:perturb-dist} shows a deformation experiment example, where Figs. \ref{fig:perturb-dist} (a) and (b) show $D_1$ and $D_2$ at step 0 respectively, computed by the Stress Majorization layout to produce stress faithful drawings. 
As the positions are perturbed in Figs. \ref{fig:perturb-dist} (c) and (d), the geometric distances between the vertices are perturbed to be more disproportionate to their graph theoretic distance.

Fig. \ref{fig:distchange_perturb_average} shows the average $DCQ$ for each deformation step, averaged for all data sets. 
$DCQ$ decreases with each deformation step, confirming Hypothesis \ref{hyp:distchangeval}.

We can also see that $DCQ_1$ decreases to a lower value in latter deformation steps compared to $DCQ_2$, which only decreases by about 0.1. 
Considering how far the drawings are from the initial distance faithful drawings at step 10, e.g. Fig. \ref{fig:perturb-dist} (d), the minor decrease with $DCQ_2$ does not capture the extent of change as closely as $DCQ_1$. 
This indicates that $DCQ_1$ is more effective at capturing the distance change faithfulness, also supporting Hypothesis \ref{hyp:distchangeval}.

\subsection{Discussion and Summary}

Our deformation experiment validates the effectiveness of $DCQ$ metrics to measure the distance change faithfulness of drawings of dynamic graphs. 
We also observe that $DCQ_1$ is more effective at capturing differences in distance change faithfulness than $DCQ_2$. 
This may be due to the fact that scaling by maximum distance in  $DCQ_2$ can be more susceptible to outliers, and may cause smaller distance changes to be underrepresented. Therefore, we will focus on $DCQ_1$ as the main comparison metric for the next experiments.

\textit{In summary, our experiments have validated Hypothesis \ref{hyp:distchangeval}, showing that $DCQ$ effectively reflects the distance change faithfulness of dynamic graph drawing, and that $DCQ_1$ captures distance change faithfulness more effectively than $DCQ_2$.}

\section{Distance Change Faithfulness Layout Comparison}

After validating the effectiveness of the distance change faithfulness metrics, we compare the performance of a number of graph drawing algorithms using the $DCQ$ metrics. We select the following layout algorithms: \textit{Stress-based} layouts \textit{Stress Majorization}~\cite{gansner2004graph} and \textit{Sparse Stress Minimization}~\cite{ortmann2016sparse}; \textit{MDS (Multi-Dimensional Scaling)} layouts \textit{Pivot MDS}~\cite{brandes2006eigensolver} and \textit{Metric MDS}~\cite{torgerson1952multidimensional}; \textit{tsNET}~\cite{kruiger2017tsnet}; \textit{FR (Fruchterman-Reingold)}~\cite{fruchterman1991graph}; and \textit{LinLog}~\cite{noack2003energy}.

Stress-based layouts aim to minimize stress (i.e. high distance faithfulness), therefore we expect them to be the most distance change faithful. 
As the concept of stress was adapted from MDS, we expect that MDS layouts will also perform quite well.
Meanwhile, we expect force-directed layouts such as FR and LinLog to be less distance change faithful. We therefore formulate the following hypothesis:
\begin{hyp}\label{hyp:distchangecompmax}
    Stress Majorization and Sparse Stress Minimization obtain the highest $DCQ$ scores, while FR and LinLog obtain the lowest $DCQ$ scores.
\end{hyp}

We again use a mix of synthetic graphs and real-world graphs from the Social Evolution set~\cite{madan2011sensing}, in total fifteen sets of dynamic graphs with 20-300 vertices.


\begin{table}[t]
\centering
\caption{Layout comparison for \(tree\_100\_1\)}
\label{table:distlayoutcomp_tree_100_1}
\begin{tabular}{|c|c|c|c|c|}
\hline
$G_1$ Stress Maj. & $G_1$ S. Stress Min. & $G_1$ Pivot MDS & $G_1$ Metric MDS & $G_1$ tsNET \\ \hline
\includegraphics[height=0.2\textwidth]{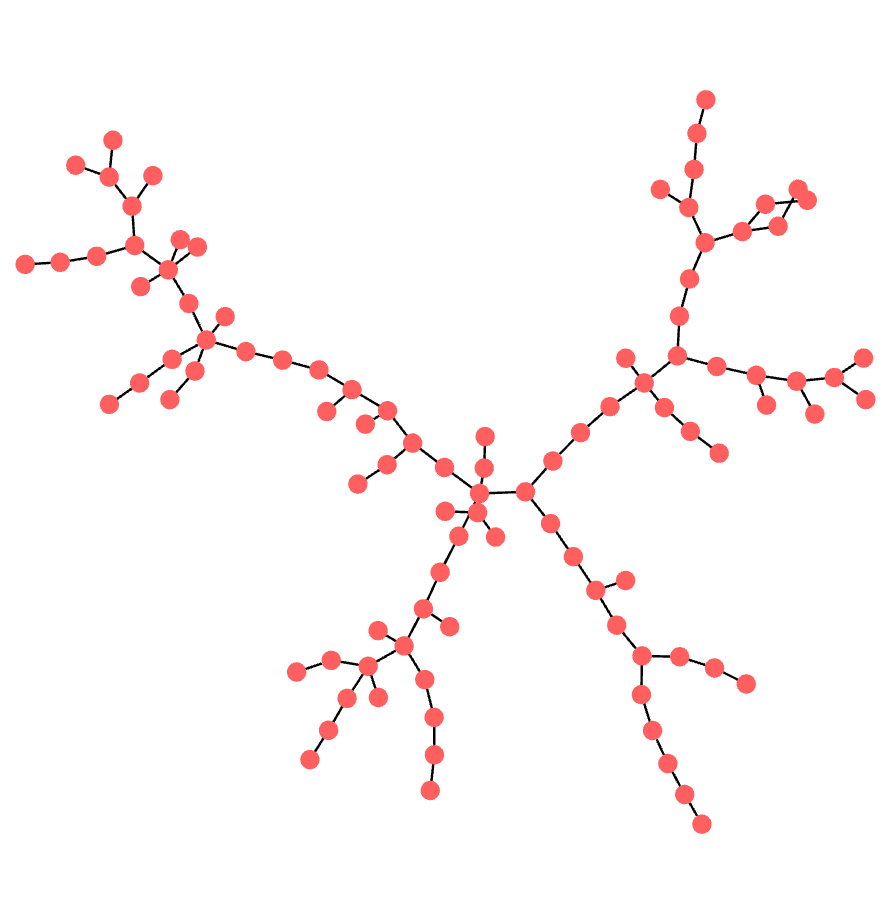} &
\includegraphics[height=0.2\textwidth]{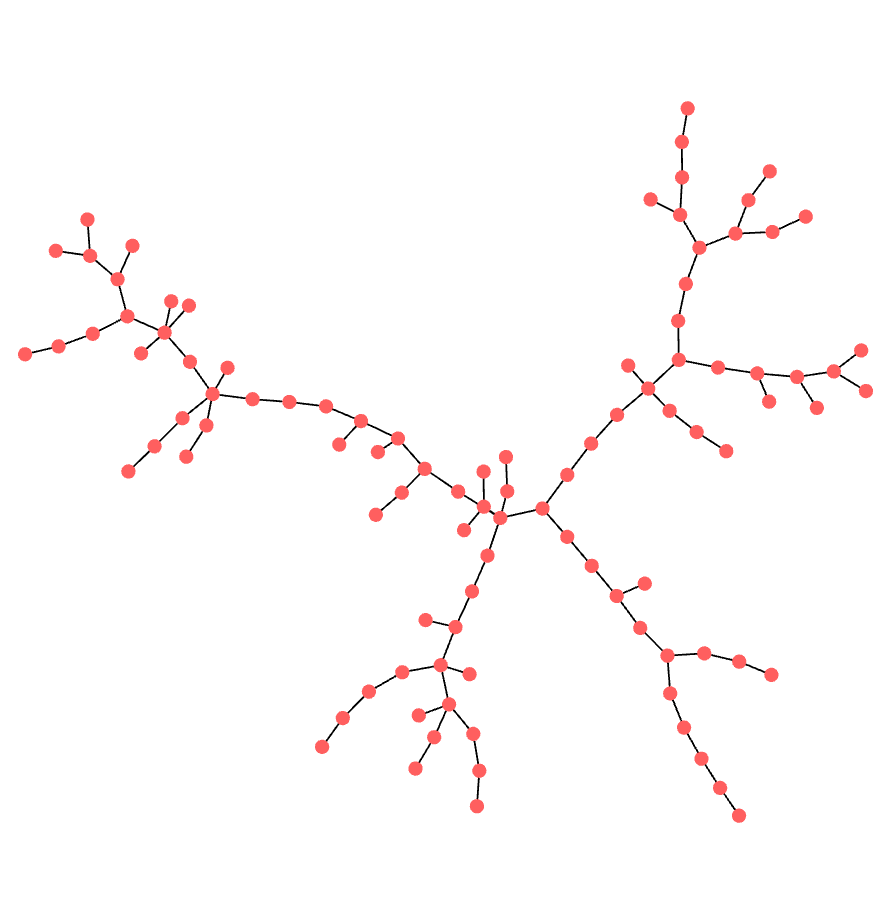} &
\includegraphics[height=0.2\textwidth]{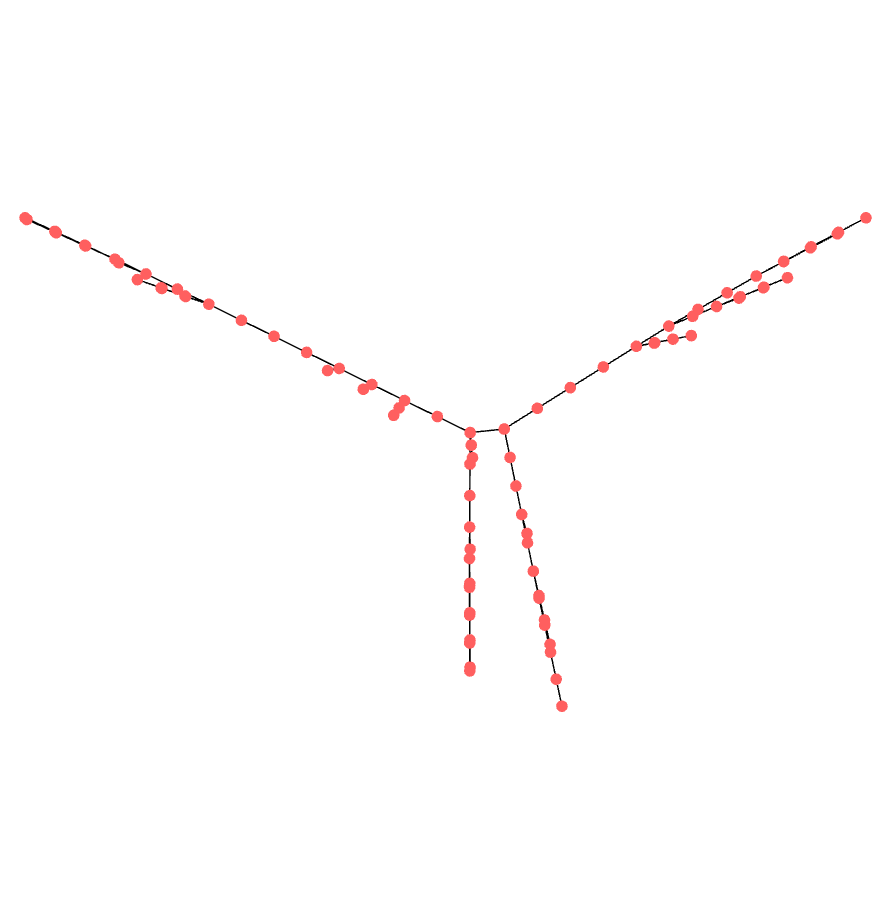} &
\includegraphics[height=0.2\textwidth]{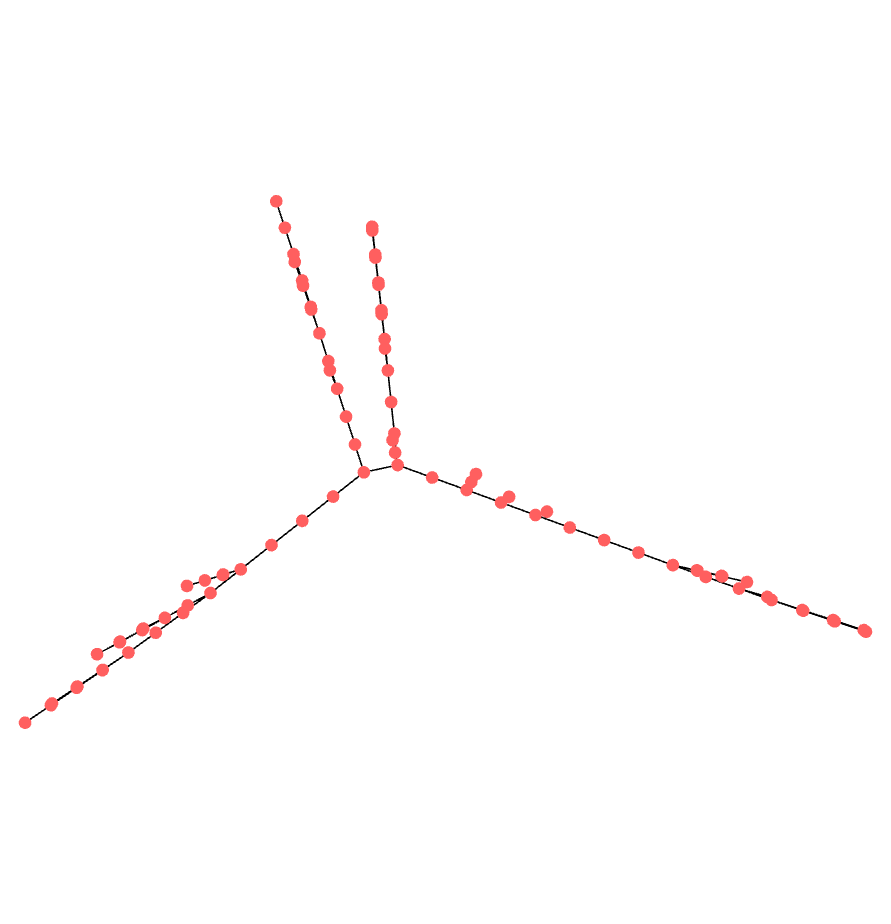} &
\includegraphics[height=0.2\textwidth]{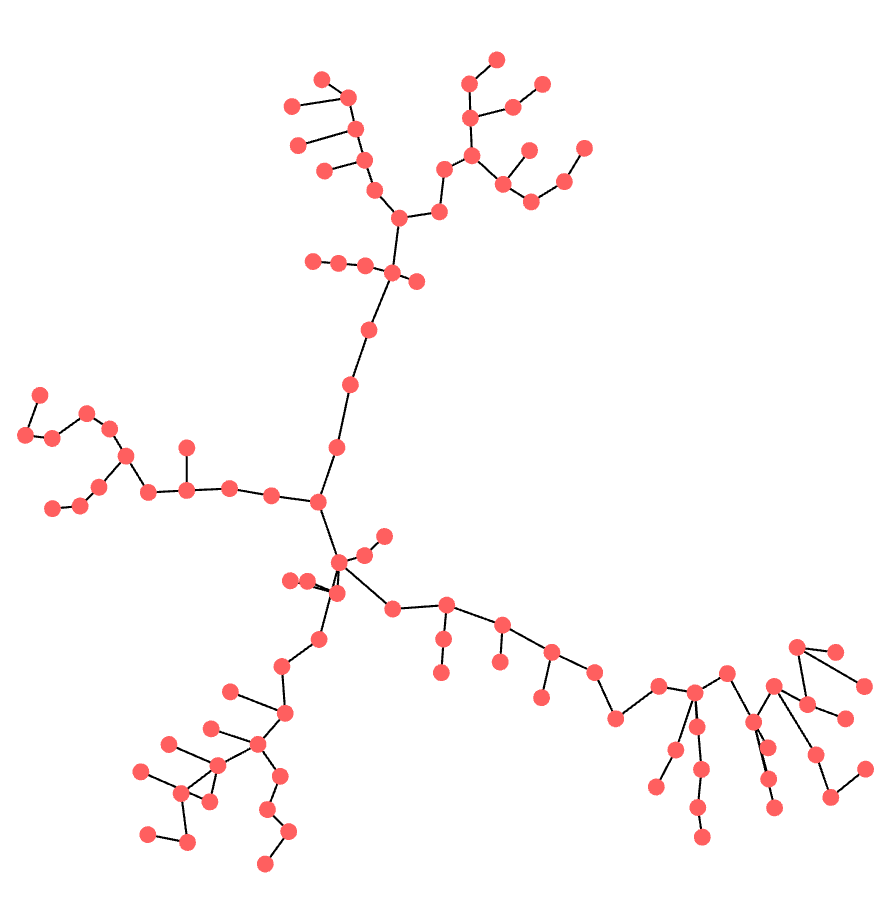}\\ \hline
$G_2$ Stress Maj. & $G_2$ S. Stress Min. & $G_2$ Pivot MDS & $G_2$ Metric MDS & $G_2$ tsNET \\ \hline
\includegraphics[height=0.2\textwidth]{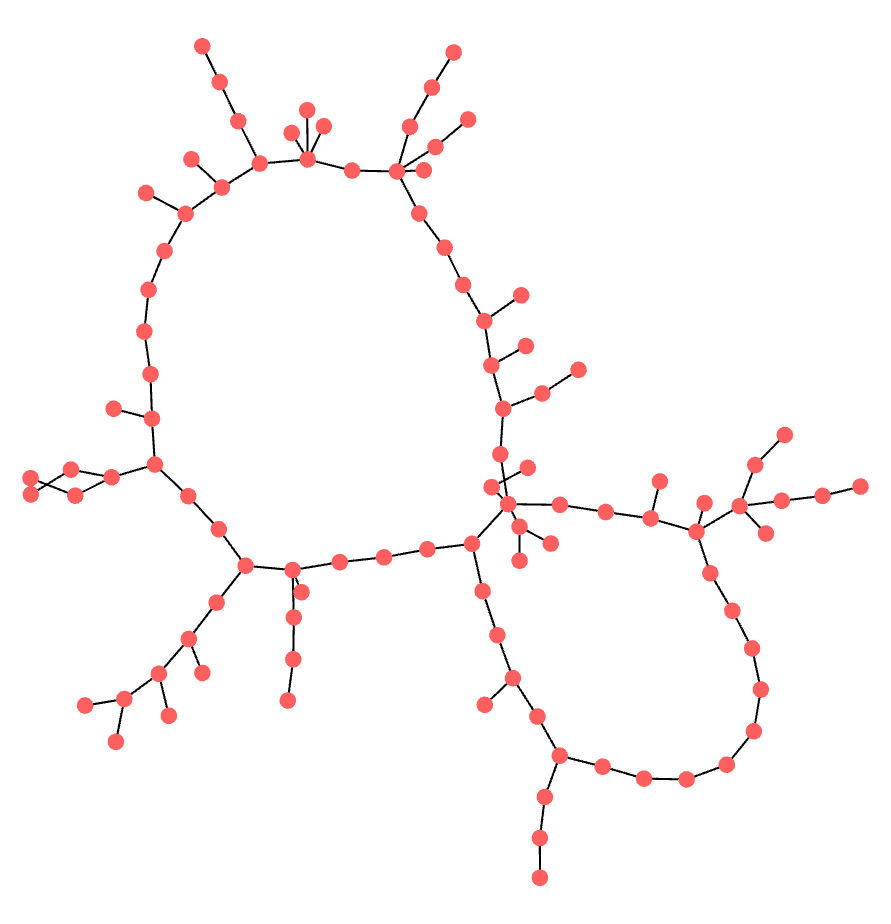} &
\includegraphics[height=0.2\textwidth]{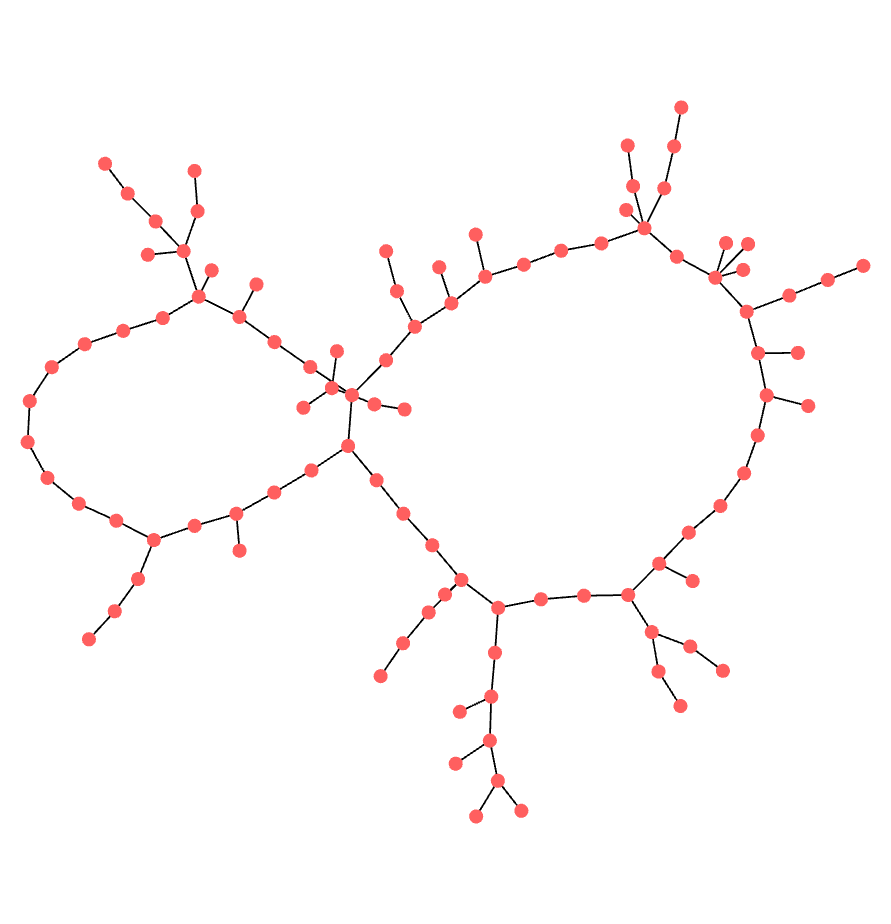} &
\includegraphics[height=0.2\textwidth]{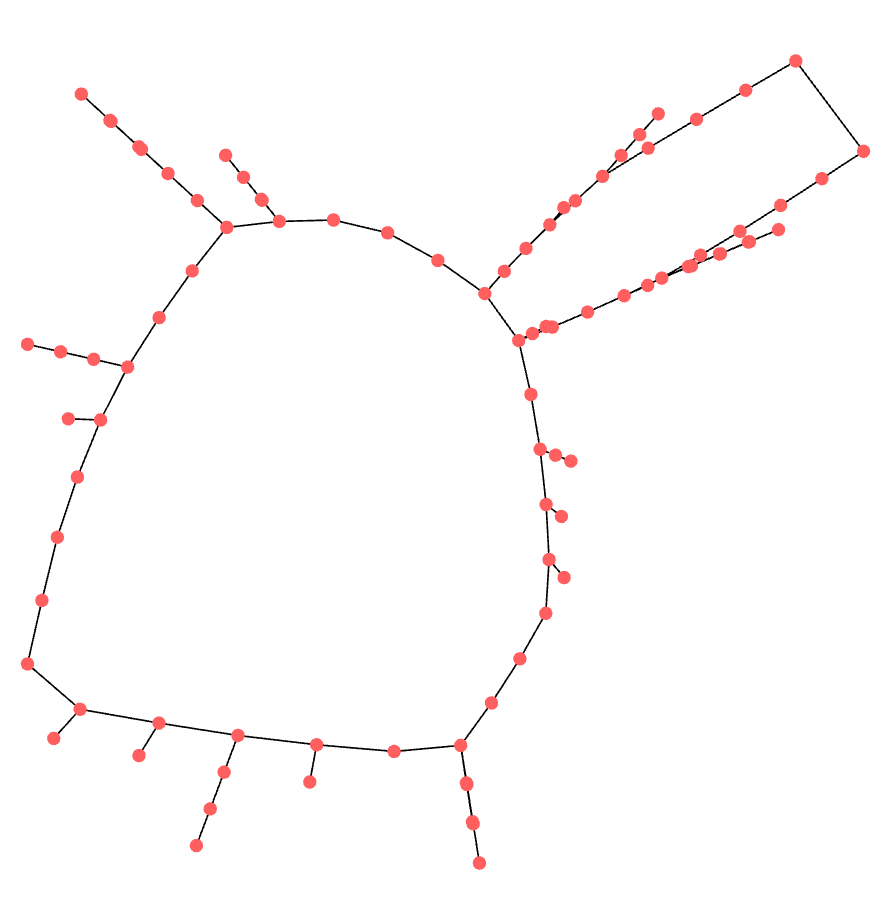} &
\includegraphics[height=0.2\textwidth]{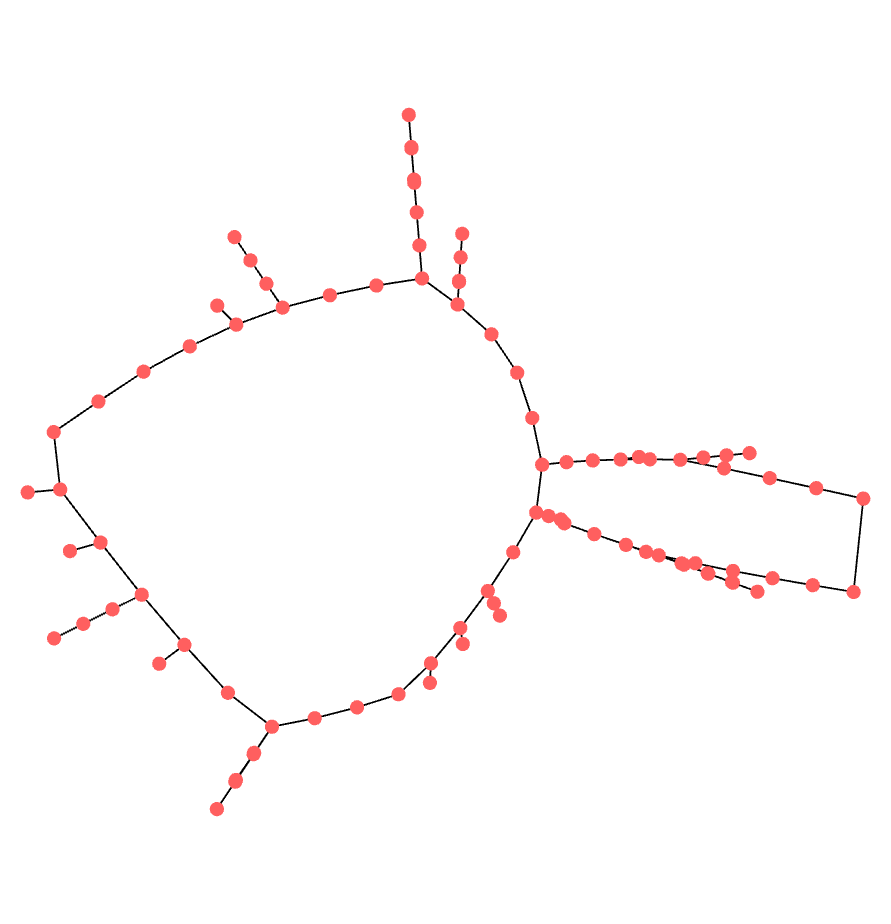} & 
\includegraphics[height=0.2\textwidth]{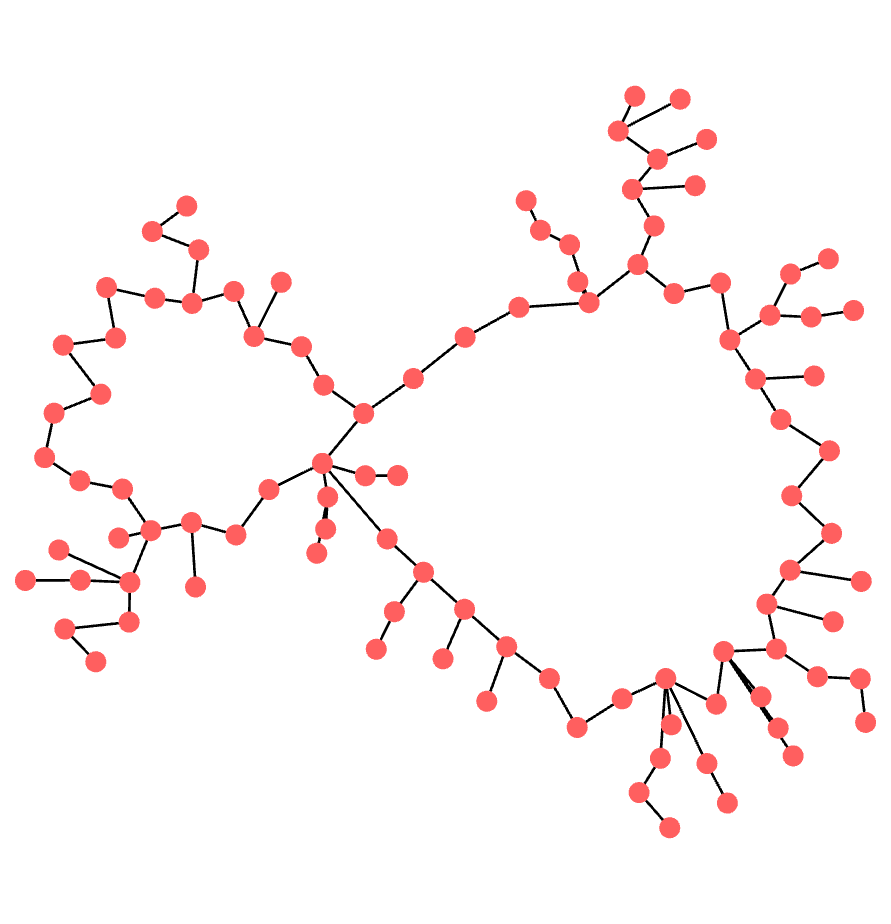}\\ \hline
$G_1$ FR & $G_1$ LinLog & \multicolumn{3}{c|}{Stress} \\ \hline
\includegraphics[height=0.2\textwidth]{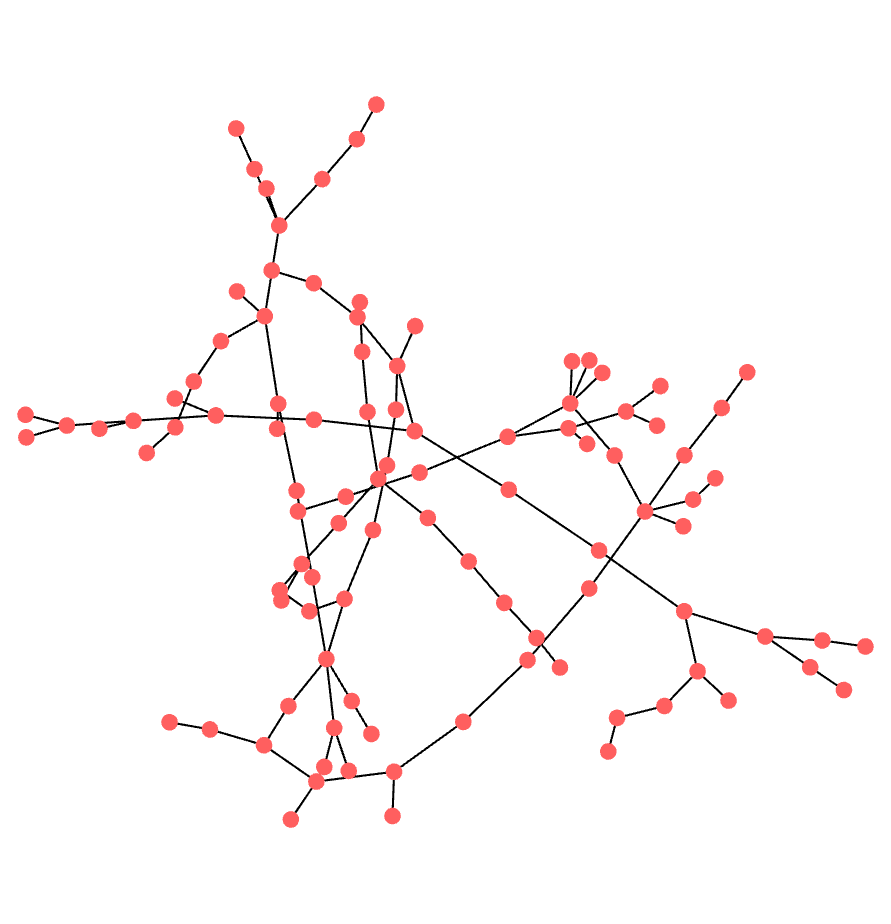} &
\includegraphics[height=0.2\textwidth]{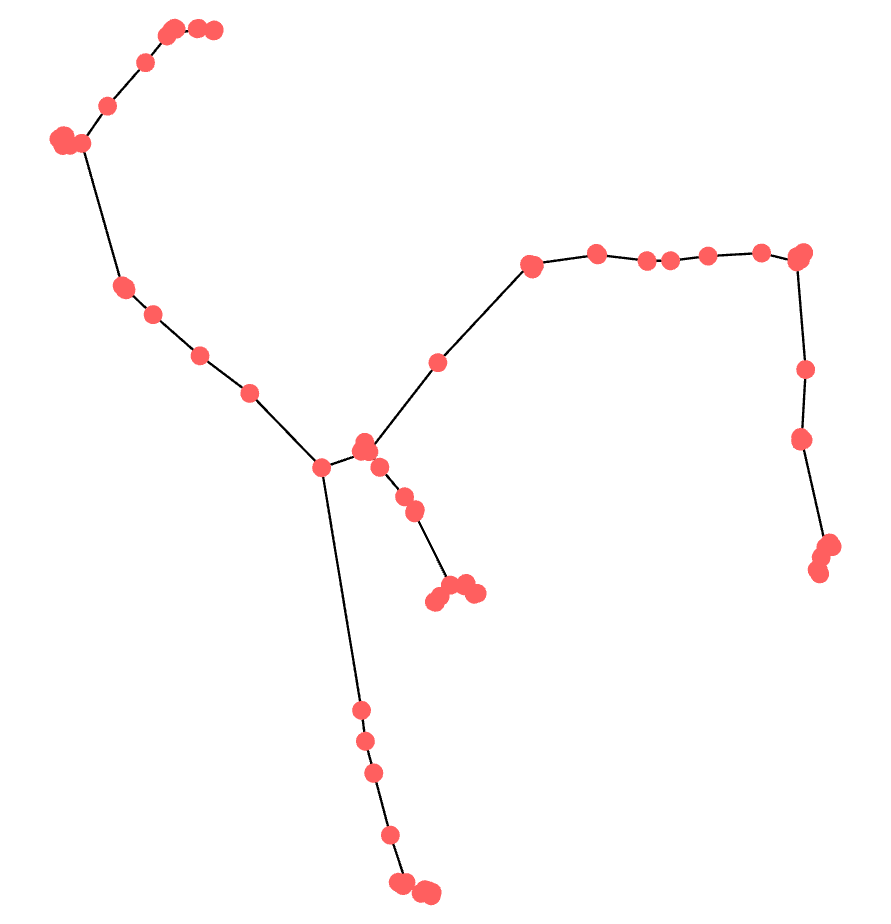} & \multicolumn{3}{c|}{\includegraphics[height=0.2\textwidth]{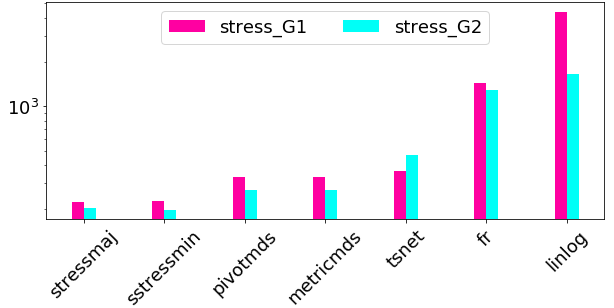}} \\ \hline
$G_2$ FR & $G_2$ LinLog & \multicolumn{3}{c|}{$DCQ$} \\ \hline
\includegraphics[height=0.2\textwidth]{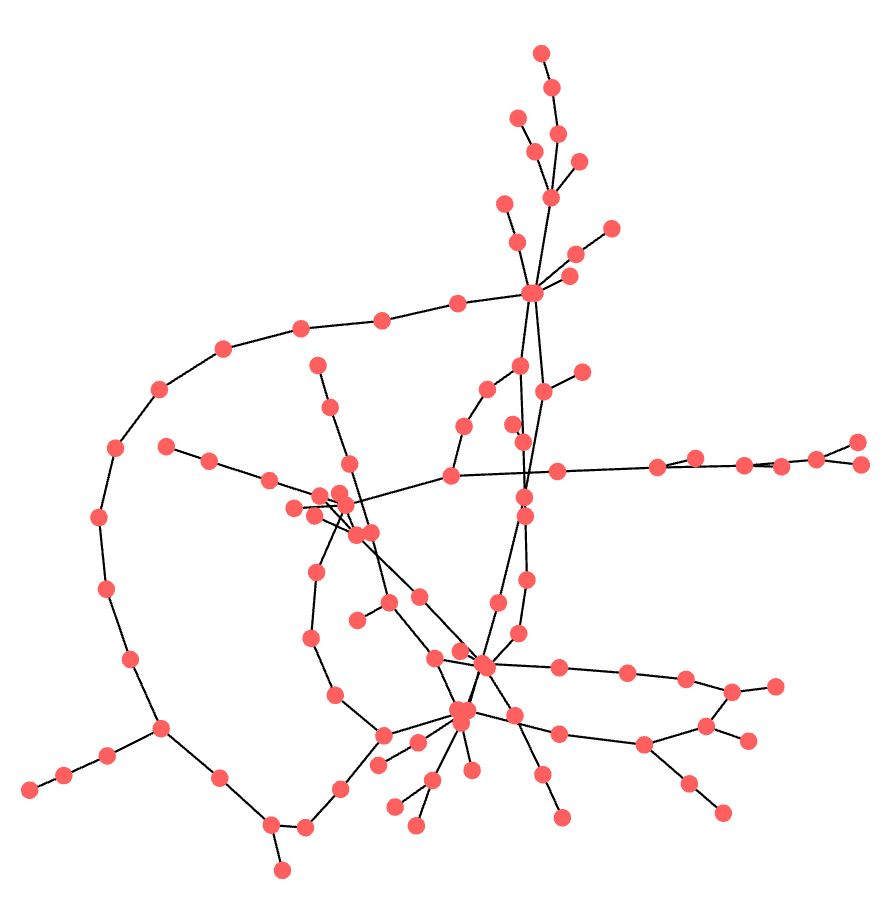} &
\includegraphics[height=0.2\textwidth]{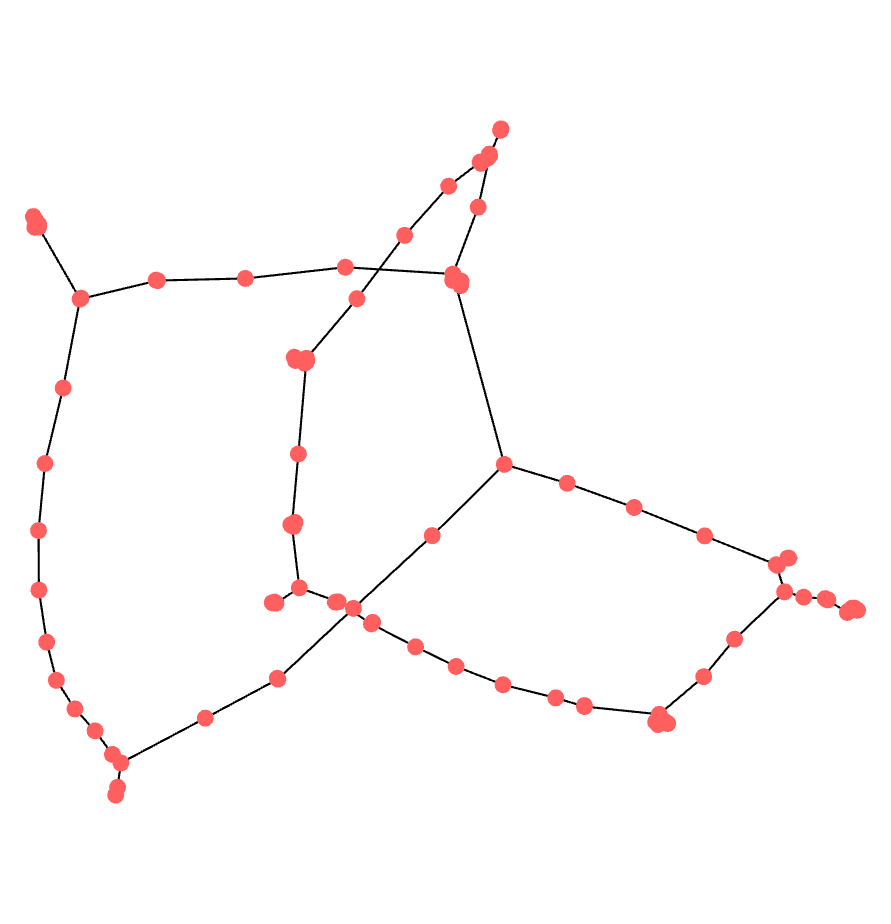} & \multicolumn{3}{c|}{\includegraphics[height=0.2\textwidth]{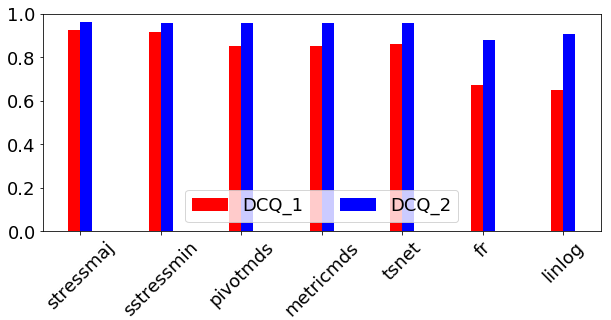}} \\ \hline
\end{tabular}
\end{table}

Table \ref{table:distlayoutcomp_tree_100_1} shows a layout comparison example.  The stress of the drawings are shown in magenta ($D_1$) and cyan ($D_2$), and $DCQ_1$ and $DCQ_2$ are shown in red and blue respectively. 
Stress Majorization and Sparse Stress Minimization obtains the two highest $DCQ$, while FR and LinLog obtain notably higher stress and lower $DCQ$ than other layouts, supporting Hypothesis ~\ref{hyp:distchangecompmax}.

\begin{figure}[t!]
\centering
\subfloat[Stress]{
\includegraphics[width=0.5\textwidth]{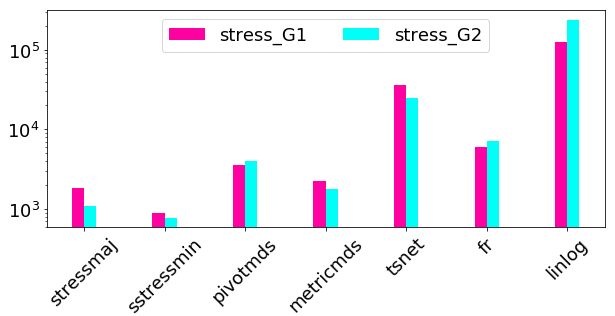}
}
\subfloat[$DCQ$]{
\includegraphics[width=0.5\textwidth]{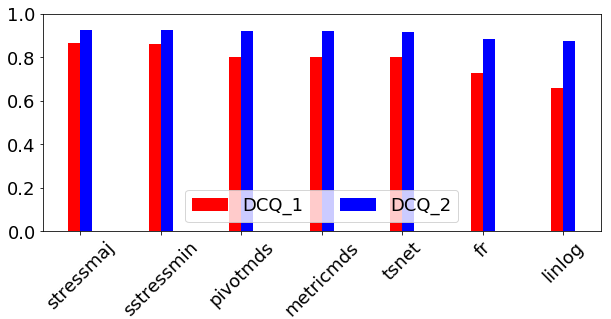}
}
\caption{(a) Average stress scores; (b) average $DCQ$ metrics for layout comparison experiments. Stress Majorization and Sparse Stress Minimization obtain the highest $DCQ$, while FR and LinLog obtain the lowest $DCQ$, supporting Hypothesis \ref{hyp:distchangecompmax}.}
\label{fig:distchange_layoutcomp_avg}
\end{figure}

Fig. \ref{fig:distchange_layoutcomp_avg} shows the average stress and $DCQ$ scores across all layout comparison experiment data sets. 
On average, Stress Majorization and Sparse Stress Minimization obtain the lowest stress and highest $DCQ$ metrics, at around 0.86 on $DCQ_1$,
and FR and LinLog obtain the highest stress and lowest $DCQ$ metric, at around 0.7 and 0.66 respectively on $DCQ_1$, supporting Hypothesis \ref{hyp:distchangecompmax}.

\subsection{Discussion and Summary}
\label{subsec:discussion}

Our experiments have supported Hypothesis \ref{hyp:distchangecompmax}, 
showing that the stress-based layouts, which explicitly aim to achieve low stress drawings, also obtain high $DCQ$, while FR and LinLog, which are not specifically designed to minimize stress, obtain lower $DCQ$.

While LinLog obtains the best results in the $CCQ$ layout comparison, in this case, it obtains the lowest $DCQ$. 
This shows a case where a layout that is optimal for one metric may not perform as well on other metrics. 

\textit{In summary, our experiments have supported Hypothesis \ref{hyp:distchangecompmax} for stress-based layouts, which obtain the highest $DCQ$ metrics on average.
We also observe that a layout obtaining good performance on one change faithfulness metric may not perform as well on other change faithfulness metrics.}

\section{Conclusion and Future Work}
\label{sec:conclusion}

We introduce a general framework for measuring change faithfulness in  dynamic graph drawings. 
Based on the framework, we present  cluster change faithfulness metrics $CCQ$ and distance change faithfulness metrics $DCQ$, as specific instances of the framework.

We validate the effectiveness of both metrics using deformation experiments,
and then compare various graph drawing layouts using the metrics. 
Our experiments confirm that LinLog obtains the highest cluster change faithfulness, while stress-based layouts obtain the highest distance change faithfulness.

Future work include designing other specific instances of the change faithfulness metric framework. 
More specifically, $DCQ$ can be extended by using other notions of distance. As the general nature of the change faithfulness metric framework allows for the development of other specific metrics, this also presents the opportunity for designing new layout algorithms to optimize such new metrics.

\newpage

\bibliographystyle{splncs04}
\bibliography{clustering_GD}

\newpage

\begin{subappendices}
\renewcommand{\thesection}{\Alph{section}}

\section{Cluster Change Faithfulness Framework}
\label{sec:ccqframework}

\begin{figure}[h]
\centering
\includegraphics[width=0.8\textwidth]{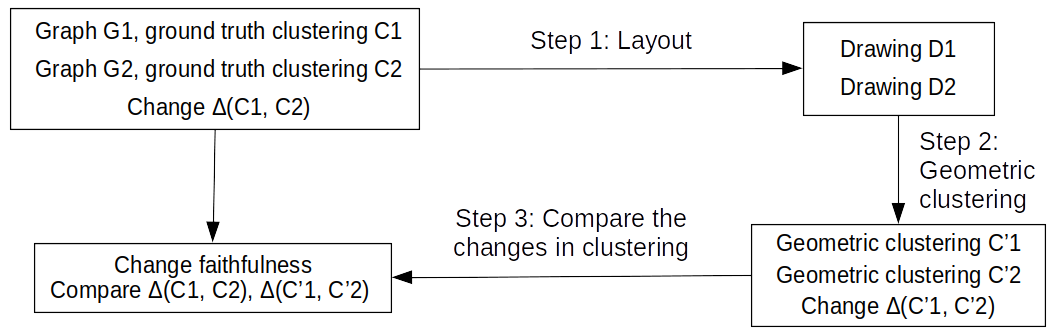}
\caption{Cluster change faithfulness metric framework: 
the cluster change faithfulness metric is computed by comparing the ground truth change in clustering $\Delta(G_1, G_2)$  and the geometric clustering change $\Delta(D_1, D_2)$.}
\label{fig:clustchangeframework}
\end{figure}

\section{Distance Change Faithfulness Framework}
\label{sec:dcqframework}

\begin{figure}[h]
\centering
\includegraphics[width=0.8\textwidth]{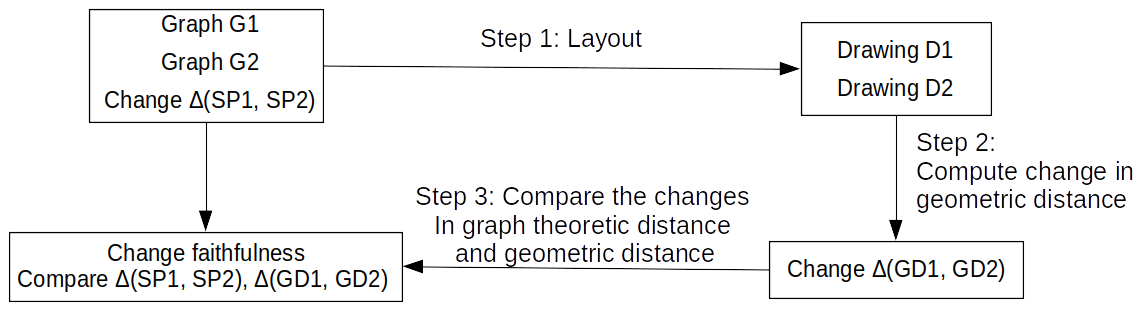}
\caption{Distance change faithfulness framework: the distance change faithfulness is computed by comparing the ground truth change in graph theoretic distance $\Delta(SP_1, SP_2)$ and the change in geometric distance in the drawing $\Delta(GD_1, GD_2)$.}
\label{fig:distchangeframework}
\end{figure}
\end{subappendices}


\end{document}